\documentclass{article}
\usepackage[utf8]{inputenc}

\usepackage{graphicx,color}

\usepackage{jcappub} 
\usepackage{url}
\usepackage[T1]{fontenc} 
\usepackage{amssymb}
\usepackage{amsmath}              
\usepackage{bm}
\usepackage{colortbl}
\usepackage{longtable}
\usepackage{subfigure}
\usepackage{multirow,comment}
\usepackage[T1]{fontenc}

\newcommand{\trh}{T_{\mathrm{RH}}}
\newcommand{\tmax}{T_{\mathrm{max}}}

\begin{document}

\begin{flushright}

UMN-TH-4129/22, FTPI-MINN-22/20 \\
August 2022
\end{flushright}

\title{Post-Inflationary Dark Matter Bremsstrahlung}
\author[a]{Yann Mambrini}
\author[b]{Keith A. Olive}
\author[c]{Jiaming Zheng}

\affiliation[a]{Universit\'e Paris-Saclay, CNRS/IN2P3, IJCLab, 91405 Orsay, France}
\affiliation[b]{William I. Fine Theoretical Physics Institute, School of Physics and Astronomy, University of Minnesota, Minneapolis, MN 55455, U.S.A.}
\affiliation[c]{Tsung-Dao Lee Institute $\&$ School of Physics and Astronomy, Shanghai Jiao Tong University, Shanghai 200240, China}

\abstract{
Dark matter may only interact with the visible sector 
efficiently at energy scales above the inflaton mass, such as the Planck scale or the grand unification scale.
In such a scenario, the dark matter is mainly produced out of equilibrium during the period of reheating, often referred to as UV freeze-in. 
We evaluate the abundance of the dark matter generated from bremsstrahlung off the inflaton decay products assuming no direct coupling between the inflaton and the dark matter. 
This process generally dominates the production of dark matter for low reheating temperatures where the production through the annihilations of particle in the thermal plasma becomes inefficient. 
We find that the 
bremsstrahlung process dominates 
for reheating temperatures $\trh \lesssim 10^{10}$ GeV, 
and produces the requisite density of dark matter for a UV scale $\simeq 10^{16}$ GeV.
As examples, we calculate numerically the yield of the dark matter bremsstrahlung through gravitation and dimension-6 vector portal effective interactions.
}

\maketitle

\section{Introduction}

Dark Matter (DM) is only indirectly observed by its gravitational effects and its nature remains a mystery. A wide variety of well-motivated extensions to the Standard Model contain stable fundamental particles as candidates for DM and may link it to other profound physics questions such as the strong CP problem, the hierarchy problem, charge quantization and gauge coupling unification. Direct searches \cite{LZ,PandaX-4T:2021bab,XENON:2018voc,LUX:2016ggv,DarkSide:2018bpj,CRESST:2019jnq}   for particle DM have excluded a large portion of the parameter space of the weakly interactive massive particle (WIMP) scenario\,\cite{Srednicki:1988ce} where the DM was in chemical equilibrium with the Standard Model (SM) sector in the early universe (see
\cite{Arcadi:2017kky,book} for reviews on the subject).

Dark matter can also be produced out of chemical
equilibrium\,\cite{ehnos,nos,Khlopov:1984pf,Olive:1984bi,Hall:2009bx,Mambrini:2013iaa,Bernal:2017kxu,Barman:2020plp,Barman:2021qds}, dubbed "freeze-in" production, if it only directly interacts with particles in the thermal plasma through a very weak interaction, even gravity \cite{Tang:2017hvq,Ema:2015dka,Ema:2016hlw,Ema:2018ucl,Ema:2019yrd,Garny:2015sjg,Garny:2017kha,Chianese:2020yjo,Chianese:2020khl,Mambrini:2021zpp,Clery:2021bwz,Clery:2022wib,Barman:2021ugy,Haque:2021mab,Aoki:2022dzd}. This scenario can also appear in grand unified
theories (GUTs) where the DM particles only interact with the visible
sector through a heavy mediator at the GUT scale 
\,\cite{Mambrini:2013iaa,Mambrini:2015vna,Nagata:2015dma,Bhattacharyya:2018evo} or an intermediate scale associated with 
moduli fields\,\cite{Chowdhury:2018tzw}, the
seesaw mechanism \cite{Dudas:2014bca,Dudas:2020sbq} or even the inflaton sector \cite{Heurtier:2019eou,Heurtier:2017nwl}.  In these scenarios, the dark matter is most efficiently generated at higher temperatures because these interactions are effectively irrelevant operators at low energy. This kind of DM production is usually called ``UV freeze-in'' in the literature.

\begin{figure}[ht!]
	\centering
	\includegraphics[width=0.95\textwidth]{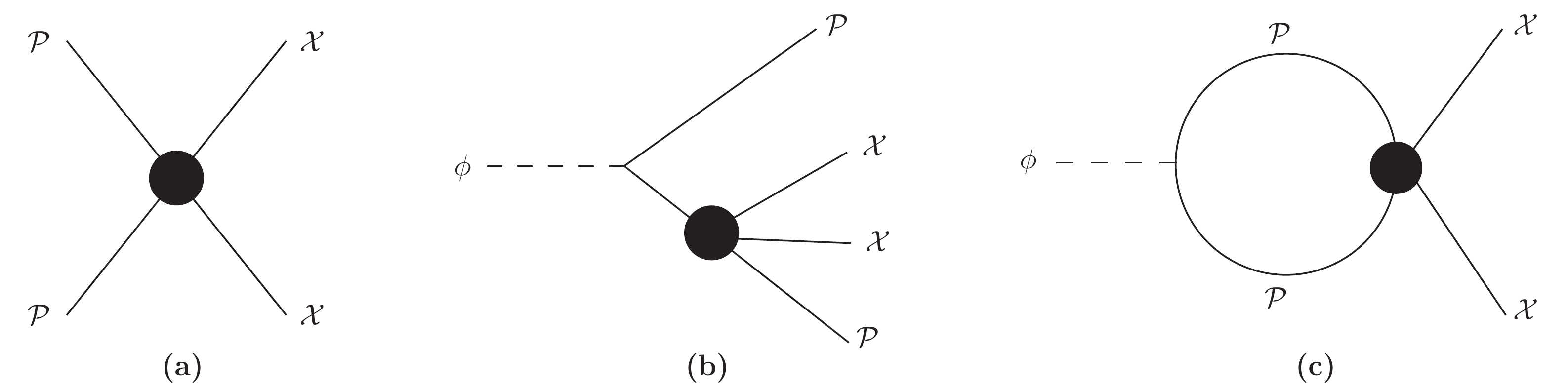}\caption{\label{fig:DM_production}DM production processes. The interaction (black bulb) that induces the annihilation of a pair of plasma particle ${\cal P}$ to a pair of dark matter ${\cal X}$ in panel (a) unavoidably	induces DM bremsstrahlung from inflaton decay in panel (b). Depend on the form of the interaction, the loop process in panel (c) may also be induced. 
	}
\end{figure}

The DM\,-\,plasma particle interactions directly induce the annihilation of plasma particles $\cal P$ into DM particles $\cal X$ as illustrated in Fig.~\ref{fig:DM_production}(a) and has been
widely considered in the literature. In a generic inflationary cosmology, the plasma particle itself is produced by the inflaton during reheating. The ${\cal P} {\cal P}-{\cal X} {\cal X}$ interaction unavoidably induces radiative dark matter production processes during inflaton decay or scattering, even
in the absence of a direct coupling of the inflaton to the dark matter
sector as we show in Fig. \ref{fig:DM_production}. 
For example, the authors of \cite{Kaneta:2019zgw} calculated the decay of the inflaton to dark matter through loop processes as in Fig.~\ref{fig:DM_production}(c) and found that it can be sufficient to generate the right relic abundance.
However, the loop process is not
always operative for every form of DM-plasma particle interaction. In particular, the loop process vanishes if the DM-plasma interaction is mediated by a heavy vector boson or by a spin-$2$ particle such as the graviton. 
 However, in addition to the loop process, there is the possibility of a four-body final state contribution to inflaton decay as shown in Fig.~\ref{fig:DM_production}(b), which is
 a bremsstrahlung-like process. As long as the interaction in (a) is present, the four-body decay
 will have a non-zero branching ratio and should be taken into
 account in the total production of dark matter. 
 
In this paper, we study the bremsstrahlung
production of dark matter during inflaton decay as in Fig.~\ref{fig:DM_production}(b). As we will show, this process tends to dominate the production of DM in a vast parameter space,
especially with relatively low reheating temperature. We will then 
focus on the comparison between Fig.~\ref{fig:DM_production}(a) and Fig.~\ref{fig:DM_production}(b), and provide numerical results for the DM bremsstrahlung rate in two UV scenarios where the loop process in Fig.~\ref{fig:DM_production}(c) vanishes, i.e., 1) when the DM only interacts gravitationally, and 2) when the DM interacts through a heavy vector boson. 

The organization of this work is as follows. In Section~\ref{sec:general consideration}, we provide general qualitative estimates of the DM bremsstrahlung. In Section~\ref{sec:GR_production} we consider the minimal 
(unavoidable) extension, where the only interaction between the visible and
dark sector is through gravity. Finally, we analyze
in Section~\ref{sec:Vector_Portal} the DM bremsstrahlung
production when the DM-plasma interaction is induced by heavy mediators, and comment briefly on possible UV completions. We conclude in Section~\ref{sec:conclusion}.

\section{Bremsstrahlung production versus plasma annihilation: general considerations
	\label{sec:general consideration}}

We start with a general estimate of the DM bremsstrahlung
during the reheating process and compare it with the production of DM through the 
scattering of particles in the plasma. 
We consider a minimal scenario where the reheating is
induced by the decay of the inflaton $\phi$
 into two particles ${\cal P}$ that rapidly thermalizes
in a hot plasma\footnote{We discuss the production of dark matter during the thermalization phase in the Appendix.}. The plasma particles ${\cal P}$ 
can interact with
the dark matter particles ${\cal X}$ through an effective interaction with energy
scale $\Lambda$. We restrict our discussion to $\Lambda\gg m_{\phi}$, so
that the production rate is not sensitive to the production threshold of the mediator. Note that $\cal P$ and $\cal X$ are symbols that represent general plasma (SM) and dark matter particles respectively disregarding their detailed properties. The ${\cal P}-{\cal X}$ interaction induces an out-of-equilibrium annihilation process
${\cal PP}\rightarrow{\cal XX}$, as shown in Fig.~\ref{fig:DM_production}(a). 
The same effective interaction also induces bremsstrahlung production
of an ${\cal X}$ pair along with ${\cal P}$ final states during
the reheating process through the inflaton
decay $\phi \rightarrow {\cal P}{\cal P}\rightarrow{\cal PPXX}$,
as shown in Fig.~\ref{fig:DM_production}(b).
We will refer to this process as the bremsstrahlung production
throughout our work. 
It turns out that the bremsstrahlung production of dark matter generally dominates when the inflaton decay rate is slow relative to the expansion rate
and results in a low reheating temperature, as we will now show. 

For concreteness, we consider a simplified scenario of reheating
where the real scalar inflaton $\phi$ mainly decays to a pair of
complex scalars $s$ or a pair of chiral fermions $f$ 
during reheating,%
\footnote{The chirality of $f$ suppresses $\phi\rightarrow f\bar{f}$. }
\begin{equation}
\phi\rightarrow s^{*}s,ff~({\rm or}~\bar{f}\bar{f})\, .
\end{equation}
From CPT, the inflaton decay rate to the $ff$ final state should be equal to 
the $\bar{f}\bar{f}$ final state. For the brevity of notation,
we will simply write $ff$ to include both channels in the rest of this
paper.  Examples of inflaton decay channels are a pair of SM
Higgs particles or a pair of right-handed neutrinos which can be easily realized by relevant operators $\phi H^\dagger H$ or $(\phi \nu_R^T C\nu_R + h.c.)$ 
 for a singlet inflaton $\phi$, where $C$ is the charge conjugate operator. We assume $m_{s},m_{f}\ll m_{\phi}$ so that the efficiency of the reheating is not blocked by phase space factors, and the inflaton decay
products thermalize the Standard Model plasma instantaneously\,\cite{Davidson:2000er,Harigaya:2013vwa,Harigaya:2014waa,Harigaya:2019tzu} (see however \cite{Garcia:2018wtq,Drees:2021lbm,Drees:2022vvn,Garcia:2021iag}).

The interactions responsible for the inflaton decays are described by the operators,
\begin{subequations}
\begin{align}
{\cal L}_{\phi s} & =\mu_{\phi s}|s|{}^{2}\phi\,,\\
{\cal L}_{\phi f} & =\frac{c_{\phi f}}{2}f^{T}Cf\phi+h.c.~\,.~\label{eq:L_reheating}
\end{align}
\end{subequations}
The corresponding inflaton decay rate to a final state
$i$ is
\begin{equation}
\Gamma_{\phi}^i=\frac{\lambda_i^{2}}{16\pi}m_{\phi}\,,
\label{eq:inflaton_decay_rate}
\end{equation}
with the effective coupling $\lambda_i$ defined as
\begin{equation}
\lambda_{s}=\frac{\mu_{\phi s}}{m_{\phi}}\,,\qquad\lambda_{f}=c_{\phi f}\,,
\label{Eq:lambdai}
\end{equation}
for $\phi\rightarrow s^{*}s$ and $\phi\rightarrow ff$, respectively.
We would expect the natural range of $\lambda_{s}$ from a UV model
to be comparable to that of $\lambda_{f}$. 

The evolution of the inflaton, the radiation(thermal plasma) and the
dark matter after the end of inflation (defined when $\ddot{a} = 0$, where $a$ is the cosmological scale factor) is governed by the system of the Friedmann and Boltzmann equations\,\cite{Chung:1998rq,Giudice:2000ex,Garcia:2017tuj},
\begin{subequations}
\begin{align}
\frac{{\rm d}\rho_{\phi}}{{\rm d}t}+3H\rho_{\phi} & =-\Gamma_{\phi}\rho_{\phi}\,,
\label{Eq:friedphi}
\\
\frac{{\rm d}\rho_{R}}{{\rm d}t}+4H\rho_{R} & =\Gamma_{\phi}\rho_{\phi}\,,\\
\frac{{\rm d}n_{{\cal X}}}{{\rm d}t}+3Hn_{{\cal X}} & =R_{{\cal X}}\,,
\label{eq:DM_prod}
\end{align}
 \end{subequations}
where $R_{{\cal X}}=R_{\phi{\cal X}}+R_{{\cal PX}}$
is the total production rate of the dark matter that includes the
contributions from the annihilation of plasma particles $R_{{\cal PX}}$
and from the DM bremsstrahlung $R_{\phi{\cal X}}$. 
Note that we have assumed
a quadratic form of the inflaton potential during 
the oscillation phase in Eq.(\ref{Eq:friedphi}). The equation of state 
and thus (\ref{Eq:friedphi}) should be modified
for more complex potentials with a minimum approximated by $V \sim \phi^k$ \cite{Garcia:2020wiy,Garcia:2020eof}.

After inflation, the universe is dominated by an oscillating inflaton field whose energy density dilutes with the scale factor approximately as $a^{-3}$. During this epoch, the
radiation temperature from the inflaton decay scales as $T\propto a^{-3/8}$ \cite{Chung:1998rq,Giudice:2000ex,Garcia:2017tuj,Garcia:2020wiy,Garcia:2020eof,Barman:2022tzk,Ahmed:2022tfm,Ahmed:2021fvt}.
If we define ${\cal Y}_{{\cal X}}\equiv n_{{\cal X}}/T^{8}$, the dark matter density during inflaton domination evolves as (see,
e.g. \cite{Mambrini:2021zpp}),
\begin{equation}
\frac{{\rm d}{\cal Y}_{\cal X}}{dT}=-\frac{8}{3}\frac{R_{\cal X}}{HT^{9}}\,,
\label{eq:DM_prod_approx}
\end{equation}
and 
\begin{equation}
H^2(T)=\frac{\rho_\phi(T)}{3 M_P^2} = 
\frac{\rho_\phi({\trh})}{3 M_P^2}\left(\frac{T}{\trh}\right)^8
=\frac{\alpha \trh^4}{3 M_P^2}\left(\frac{T}{\trh}\right)^8,
\label{Eq:ht}
\end{equation}
where $\alpha\equiv\pi^{2}g_{*}/30$ with $g_{*}$ the effective degrees of freedom for the thermal plasma density at $\trh$ and ${M}_{P}\equiv1/\sqrt{8\pi G_{N}}\simeq 2.4\times 10^{18}$ GeV
is the reduced Planck mass. 
Note that we have defined the reheating
temperature $T_{\rm RH}$ as the plasma temperature when $\rho_{\phi}(\trh)=\rho_{R}(\trh)=\rho_{\rm RH}$. With the approximate matter domination down to $\trh$, the reheating temperature can be analytically expressed as
\cite{Garcia:2017tuj,Garcia:2020wiy},
\begin{equation}
\trh \approx\left(\frac{12}{25\alpha}\right)^{1/4}\left(\Gamma_{\phi}M_{P}\right)^{1/2}\,.\label{eq:mphi_Treh}
\end{equation}
Therefore, it is sufficient to know the temperature dependence of the production rate, $R_{\cal X} (T)$ to solve equation (\ref{eq:DM_prod_approx}) which determines the dark matter abundance. The temperature dependence is different for production initiated from the thermal bath compared to the bremsstrahlung production process introduced here. The latter will dominate in different regimes depending on the value of $\trh$.

For the annihilation process ${\cal PP}\rightarrow{\cal XX}$ as in Fig.~\ref{fig:DM_production}(a), 
we express the dark matter production rate for $T\gg m_{{\cal X}}$
as, 
\begin{equation}
R_{\cal PX} = \frac{T^{6+n}}{\Lambda^{2+n}}\,,
\label{Eq:rateplasma}
\end{equation}
where $\Lambda$ represents the UV scale.
In fact, if one wants to be more precise, the interactions between
the dark matter
${\cal X}$ and the plasma particles ${\cal P}$ 
can be parametrized by 
\begin{equation}
    \frac{1}{\Lambda^{2+n}} =   \frac{{\cal O}(1)}{\pi^5} \frac{1}{M^{2+n}}\,,
    \label{eq:phasecount}
\end{equation}
where the factors
of $\pi$ are being counted 
to monitor the phase space factors of both the initial particle densities and the final states.
These can make up a few orders of magnitude and become important when comparing with DM bremsstrahlung that is further suppressed by phase space. 
This parameterization is applicable to a wide variety of 
scenarios. 
Indeed, $M$ can then directly be  
identified with the mass of a heavy mediator in $SO(10)$ \cite{Mambrini:2013iaa,Mambrini:2015vna,Bhattacharyya:2018evo}, or if
a graviton is exchanged \cite{Mambrini:2021zpp,Clery:2021bwz,Clery:2022wib}, 
$M\simeq M_P$, or related to the SUSY breaking scale in supergravity constructions \cite{Garcia:2017tuj,Dudas:2017kfz,Dudas:2018npp}
where $M \simeq \sqrt{m_{3/2}M_P}$, 
$m_{3/2}$ being the gravitino mass.

The same scale-suppressed effective interaction responsible for the annihilation production of DM inevitably induces the bremsstrahlung
$\phi\rightarrow{\cal PPXX}$ as shown in Fig.~\ref{fig:DM_production}(b).
We can estimate the branching ratio of the bremsstrahlung process by a mere dimensional analysis with phase space counting. For example, for $n=2$
\begin{align}
{\rm Br}_{{\cal X}}^{{\rm brem}} 
& =\frac{\Gamma_{\phi\rightarrow{\cal PPXX}}}
{\Gamma_{\phi}}
\equiv \tilde{c}\left(\frac{1}{4\pi^{2}}\right)^{2}
\frac{\left(m_{\phi}/4\right)^{4}}{M^{4}}\,.
\label{Eq:brbrem}
\end{align}
For other values of $n$, the branching ratio will scale as $(m_{\phi}^{2+n}/M^{2+n})$. The interaction determines the numerical factor $\tilde{c}$ and may contain an implicit dependence on $n$. We will show some typical examples of $\tilde{c}\sim {\cal O}(1)$ for $n=2$ in Sections ~\ref{sec:GR_production} and \ref{sec:Vector_Portal}. 
The expression above has adopted the parametrization in Eq.~(\ref{eq:phasecount}). 
This expression can be compared with the loop process given in Fig.~\ref{fig:DM_production}c (assuming the loop process does not vanish). The loop branching fraction was calculated in \cite{Kaneta:2019zgw} leading to
\begin{align}
{\rm Br}_{{\cal X}}^{{\rm loop}} 
& =\frac{\Gamma_{\phi\rightarrow{\cal XX}}}
{\Gamma_{\phi}}
= \left(\frac{1}{4\pi^{2}}\right)^{2} \left(1+\frac{\pi^4}{4} \right)
\frac{m_{\phi}^{4}}{M^{4}}\,.
\label{Eq:brloop}
\end{align}
for $n=2$.
In the case of bremsstrahlung we counted the phase space of each extra
particle in addition to the two body phase space as $1/(4\pi^{2})$ (these are the same factors of $\pi$ which appear in the loop calculation), the factor
$m_{\phi}/4$ reflects the typical energy of the decay product. 
As one can see the loop process is typically two orders of magnitude larger than bremsstrahlung (for ${\tilde c} \sim 1$). However as we have said repeatedly, not all couplings lead to a non-zero rate for the loop in Fig.~\ref{fig:DM_production}c.

The production rate (the number of dark matter particles produced per unit volume and unit of time) for bremsstrahlung can then be written as
\begin{equation}
R_{\phi{\cal X}}= 2\Gamma_{\phi}{\rm Br}_{\cal X}^{\rm brem}\rho_{\phi}/m_{\phi}\,.
\label{Eq:ratebrem}
\end{equation}

Dark matter can also be produced from plasma annihilation
as long as it is lighter than the maximum temperature $T_{{\rm max}}$
during the epoch of inflaton oscillations\,\cite{Chung:1998rq,Giudice:2000ex,Garcia:2017tuj}.
If we combine Eq.(\ref{eq:DM_prod_approx}), (\ref{Eq:ht}), and (\ref{Eq:rateplasma}),
we see that $d {\cal Y}_{\cal X}/dT\propto T^{n-7}$.
Thus, for $n<6$,
DM is mainly produced at lower temperature near the threshold
at $T_{c}\simeq{\rm Max}[\trh,m_{{\cal X}}]\ll T_{{\rm max}}$, while
for $n \geq 6$, the dark matter is produced more efficiently at higher temperature
up to $T_{{\rm max}}$. The estimated dark matter density at the reheating temperature $\trh$ is
obtained by integrating Eq.(\ref{eq:DM_prod_approx}) 
between $T_{\rm max}$ and $T_c$, 
with the rate given by Eq.(\ref{Eq:rateplasma}).
We obtain \cite{Kaneta:2019zgw,Garcia:2020eof},
\begin{subequations}
\begin{align}
n_{{\cal X}}^{{\rm ann}}(\trh)\simeq & \frac{8}{\sqrt{3\alpha}(6-n)}
\frac{{M}_{P}\trh^{10}}{\Lambda^{n+2}T_{c}^{6-n}}\,,\quad n<6\,,\\
n_{{\cal X}}^{{\rm ann}}(\trh)\simeq & \frac{8}{\sqrt{3\alpha}}
\frac{{M}_{P}T_{\rm RH}^{10}}{\Lambda^{8}}\ln\frac{T_{{\rm max}}}{T_c}\,,\quad n=6\,,\\
n_{{\cal X}}^{{\rm ann}}(\trh)\simeq & \frac{8}{\sqrt{3\alpha}(n-6)}
\frac{{M}_{P}\trh^{10}T_{{\rm max}}^{n-6}}{\Lambda^{n+2}}\,,\quad n>6\,.
\end{align}
\label{eq:DM_n_ann}
\end{subequations}
It should be noted that there is an ${\cal O}(1)$ correction in the exact dark matter relic
density for $n\leq6$ since the inflaton domination approximation breaks down at
$\trh$ \cite{Ellis:2015jpg,Garcia:2017tuj}. There is also a correction for the production of dark matter produced before the radiation reaches the temperature $T_{\rm max}$ \cite{Mambrini:2021zpp}.  Nevertheless, the
important feature of these solutions, 
is that for $n\geq6$, the production 
is enhanced by the maximum temperature during inflaton oscillations, i.e.,  at temperatures which can be much larger than $\trh$.

For DM bremsstrahlung, integrating Eq.~\eqref{eq:DM_prod_approx} with the rate
given by Eq.(\ref{Eq:ratebrem}) 
between $\trh$ and $T_{{\rm max}}$ 
gives 
\begin{align}
n_{{\cal X}}^{{\rm brem}}(\trh) & \simeq\frac{ 10\alpha}{3}{\rm Br}_{{\cal X}}^{{\rm brem}}\frac{\trh^{4}}{m_{\phi}}\,,\label{eq:brem_prod}
\end{align}
and
\begin{subequations}
\begin{align}
n_{{\cal X}}^{{\rm brem}}/n_{{\cal X}}^{{\rm ann}}\simeq & \frac{ 10\alpha^{3/2}(6-n) \tilde{c}\pi}{512\sqrt{3}}
\frac{\left(m_{\phi}/4\right)^{n+1}T_{c}^{6-n}}{{M}_{P}\trh^{6}}\,,\quad n<6\,,\\
n_{{\cal X}}^{{\rm brem}}/n_{{\cal X}}^{{\rm ann}}\simeq & \frac{10\alpha^{3/2} \tilde{c}\pi}{512\sqrt{3}}
\frac{\left(m_{\phi}/4\right)^{7}}{{M}_{P}\trh^{6}}\left(\ln\frac{T_{{\rm max}}}{T_{c}}\right)^{-1}\,,\quad n=6\,,\\
n_{{\cal X}}^{{\rm brem}}/n_{{\cal X}}^{{\rm ann}}\simeq & 
\frac{10\alpha^{3/2}(n-6) \tilde{c}\pi}
{512\sqrt{3}}\frac{\left(m_{\phi}/4\right)^{n+1}}{{M}_{P}\trh^{6}T_{{\rm max}}^{n-6}}\,,\quad n>6\, ,
\end{align}
\label{eq:DM_n_ratio}
\end{subequations}
assuming 2 $\chi$'s are produced in each process. 

\begin{figure}[t]
	\centering
	\includegraphics[width=0.37\textwidth]{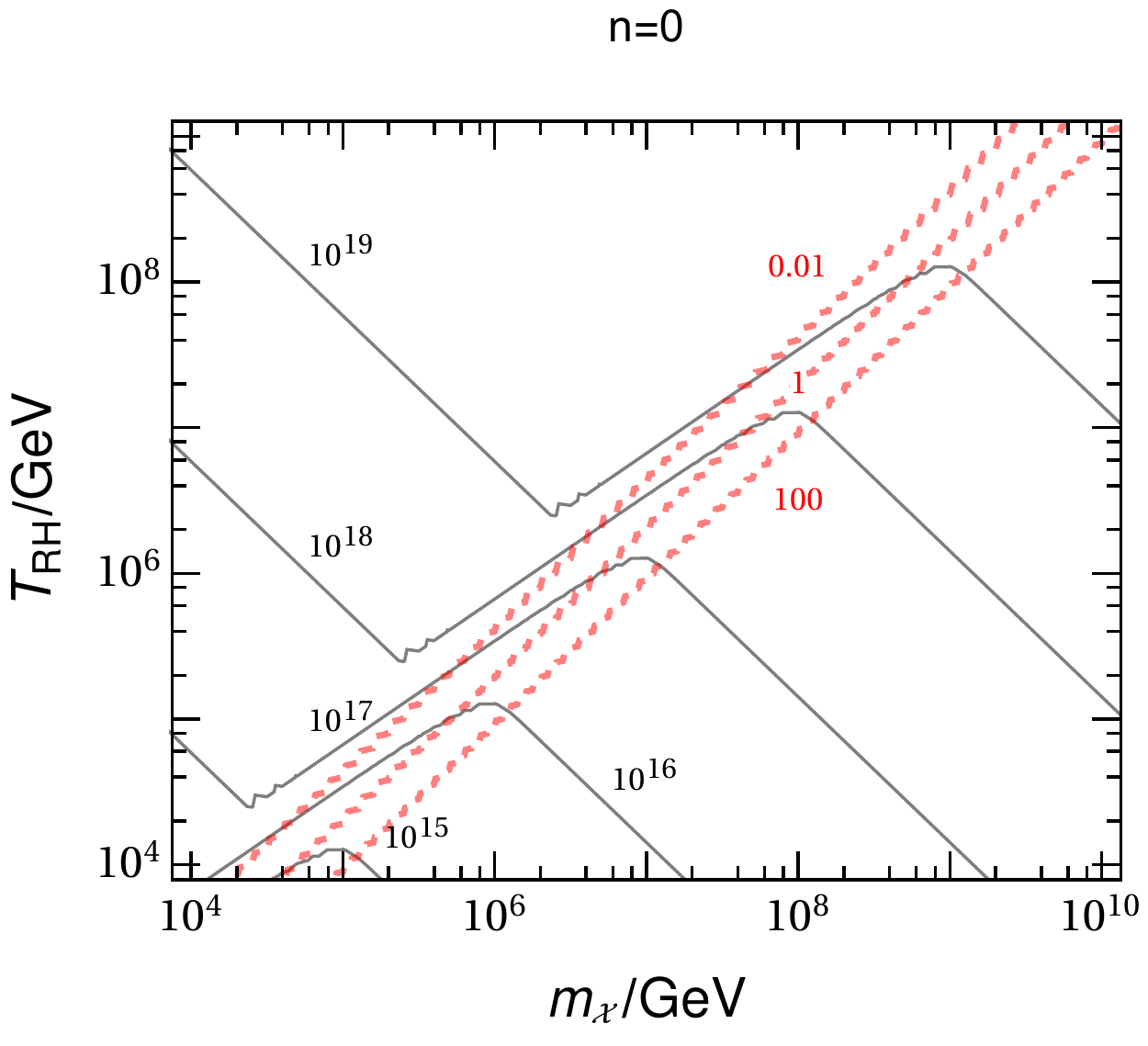}%
        \hspace*{5mm}
        \includegraphics[width=0.37\textwidth]{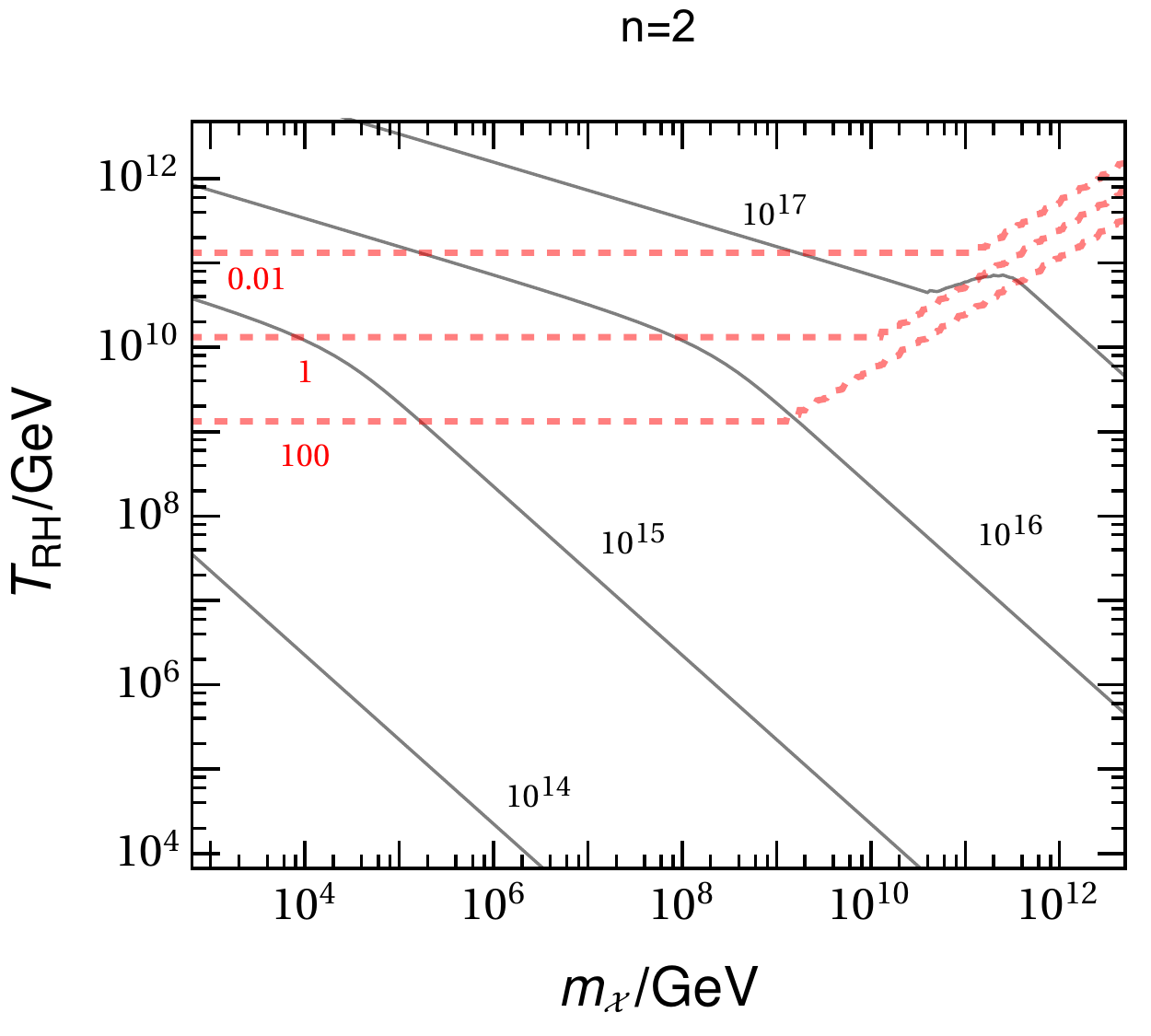}%
        \\
	\includegraphics[width=0.37\textwidth]{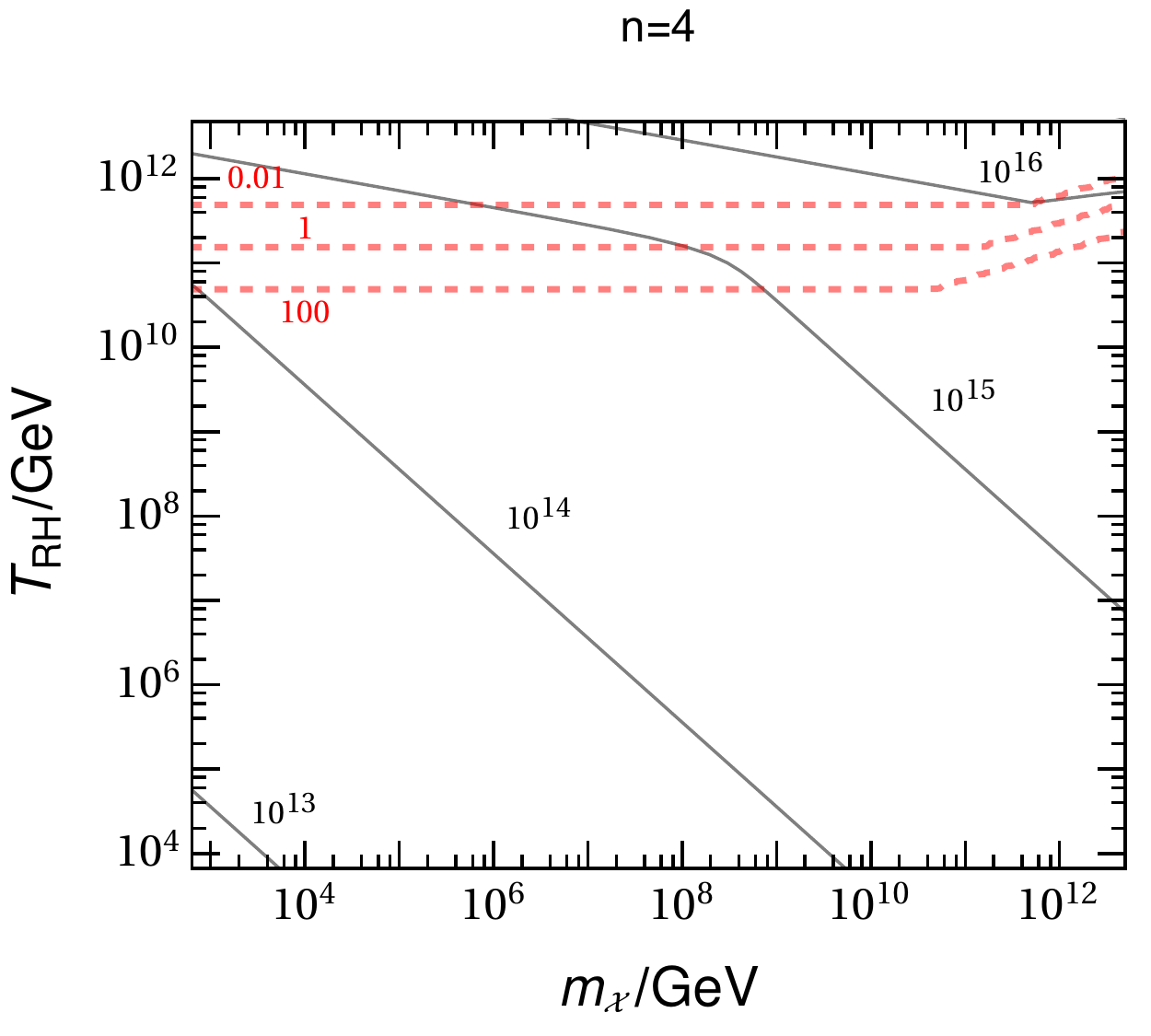}%
        \hspace*{5mm}
        \includegraphics[width=0.37\textwidth]{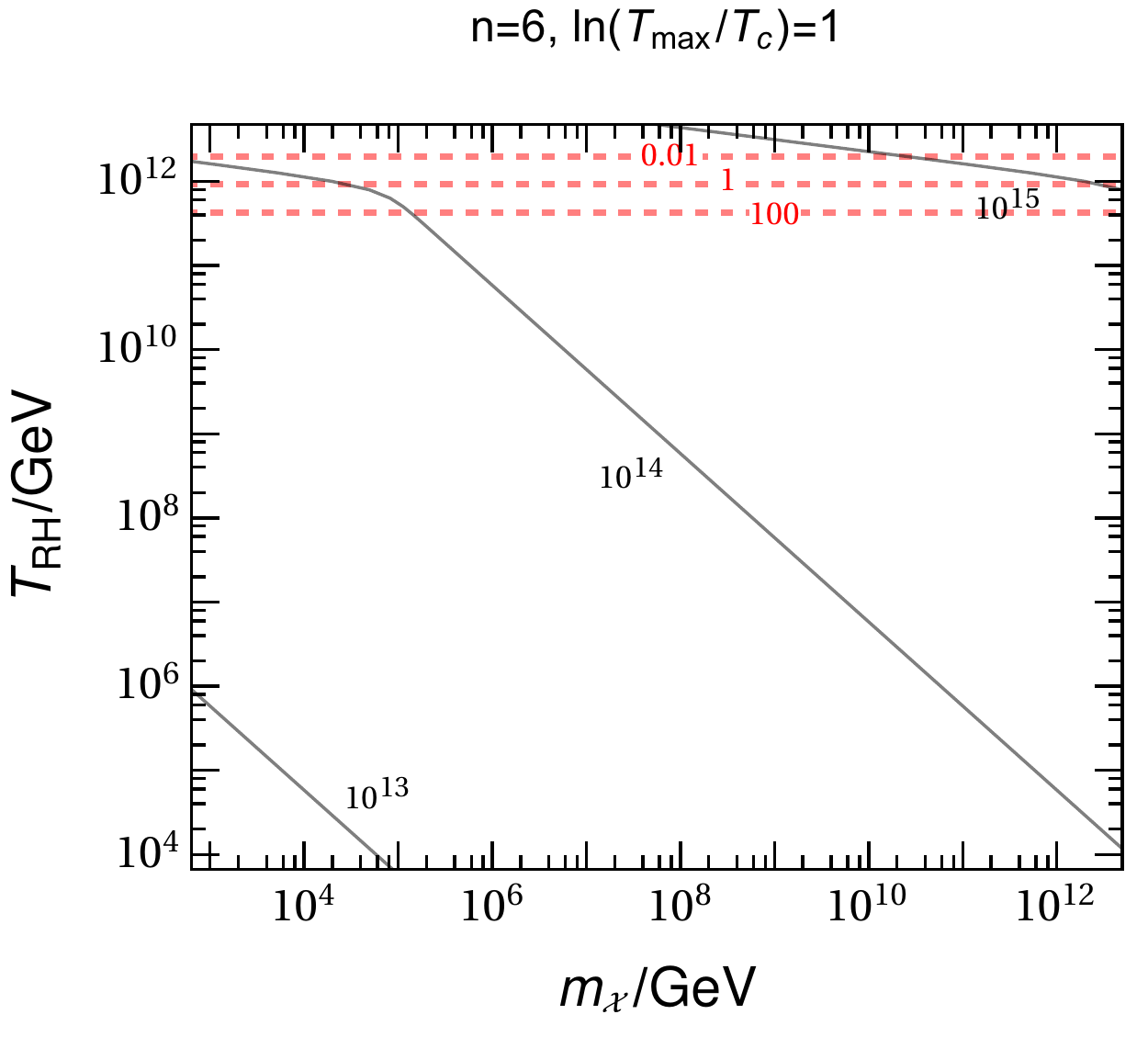}%
\caption{\label{fig:gen_compare} Parameter values that achieve the observed
		dark matter relic density \cite{Planck:2018vyg} for $n=0$, $n=2$, $n=4$, and $n=6$ with $m_{\phi}=10^{13}$GeV, $g_* =100$ and $\tilde{c}=1$. The black contours show the required
		parameter $\Lambda$ in GeV; The red line presents the
		relative importance $n_{{\cal X}}^{{\rm brem}}/n_{{\cal X}}^{{\rm ann}}$
		of bremsstrahlung vs. plasma annihilation production for the
		choice of $\Lambda$ that achieves a realistic DM relic density. }
\end{figure}

In Fig.~\ref{fig:gen_compare}, we show the parameter space which achieves the observed dark matter relic density inferred from CMB observations \cite{Planck:2018vyg} for 
$n=0$, $n=2$, $n=4$ and $n=6$ with $m_{\phi}=10^{13}$GeV, $g_*=100$ and $\tilde{c}=1$. For $n=6$ we set $\ln(T_{\rm max}/T_c)=1$ for illustration. The black contours show the required value of $\Lambda$ in GeV for each choice of $m_\chi$ and $\trh$. The red-dashed lines show the relative importance of the bremsstrahlung contribution relative to production via plasma annihilation, $n_{{\cal X}}^{{\rm brem}}/n_{{\cal X}}^{{\rm ann}}$, as given in Eq.~(\ref{eq:DM_n_ratio}). 
For $n=0$, the ratio $n_{{\cal X}}^{{\rm brem}}/n_{{\cal X}}^{{\rm ann}}\simeq 4 \times 10^{-5} (T_c/\trh)^6$ and thus
the bremsstrahlung process only dominates when $m_\chi > \trh$. %
\footnote{The tiny wiggles near the first turning point in the black contours in Fig.~\ref{fig:gen_compare}a are artifacts of the limited numerical resolution of the plot.}

For larger $n$,
it is interesting to note that the bremsstrahlung process dominates when $m_\chi < \trh$, for sufficiently low $\trh$. For example, for $n=2$,
for reheating temperatures 
below $\trh \lesssim 10^{10}$ GeV (with $m_\chi < 10^{10}$ GeV), the bremsstrahlung production process, 
which to date has never been taken into account in  the literature, 
is the dominant process which populates the dark matter in the Universe. 
Indeed, taking a dimension-six effective interaction ($n=2$) between DM and the plasma (SM) and $m_{{\cal X}}\ll \trh$, one finds for example
\begin{equation}
\frac{n_{{\cal X}}^{{\rm brem}}}{n_{{\cal X}}^{{\rm ann}}}\simeq {1.7}\, \tilde{c}\times\left(\frac{g_{*}}{100}\right)^{3/2}\left(\frac{10^{10}\,{\rm GeV}}{\trh}\right)^{2}\left(\frac{m_{\phi}}{10^{13}{\rm GeV}}\right)^{3}\,.~\label{eq:brem_vs_ann}
\end{equation}
For $m_{\phi}=10^{13}\,$GeV, the bremsstrahlung processes dominate
for low reheating temperature $\trh\lesssim10^{10}$ GeV corresponding to
$\lambda_i\lesssim 4 \times 10^{-5}$, and agrees with
our numerical result showed in Fig.~\ref{fig:gen_compare}.
This is one of the main results of our work.
Moreover, we show in the Appendix that this result
is quite robust, even when taking into account
the possibility for production through annihilation during the thermalization phase of reheating \cite{Garcia:2018wtq,Harigaya:2019tzu}.

We can also observe a change in the slope of $\trh$ vs. $m_\chi$ for fixed $\Lambda$, when the black curves cross $n_{{\cal X}}^{{\rm brem}}/n_{{\cal X}}^{{\rm ann}} \approx 1$ as shown by the red lines. This is due to the strong
difference in the dependence on $\trh$ for the two processes.
Whereas the bremsstrahlung process
can be seen as a mere decay process (rescaled by a constant branching fraction), which implies
 $n_{\cal X}^{\rm brem} \propto \trh^4$, 
the annihilation 
channels has a stronger dependence on the temperature due to its origin, proportional to the density squared of the particles 
in the bath, thus $n_{\cal X}^{\rm ann}\propto \trh^{10}/T_c^{6-n}$,  see Eqs.(\ref{eq:DM_n_ann}). For $m_{\cal X} < \trh$, $T_c = \trh$, and $n_{\cal X}^{\rm ann}\propto \trh^{4+n}$. Otherwise, $n_{\cal X}^{\rm ann}\propto \trh^{10}/m_{\cal X}^{6-n}$.
In the specific case of $n=0$, 
the slope for $\trh$ vs. $m_{\cal X}$ for the bremsstrahlung and 
annihilation processes are the same. This is because annihilations dominate when $m_{\cal X} < \trh$, and $n_{\cal X}^{\rm ann}\propto \trh^{4}$ as it does for the bremsstahlund process which dominates at higher masses.  This is clearly seen in the upper left panel of 
Fig.~\ref{fig:gen_compare}.
For $n\neq0$, the density from annihilations has a steeper dependence on the temperature. 
The relic density today as a fraction of critical density, $\Omega_{\cal X} h^2 \propto n_{\cal X} m_{\cal X}/ \trh^3$ and
we obtain $\Omega^{\rm brem}_{\cal X} h^2 \propto m_{\cal X} \trh$ for the
bremsstrahlung production, and $\Omega^{\rm ann}_{\cal X} h^2 \propto m_{\cal X} \trh^{1+n}$ for production from the thermal plasma (when $m_{\cal X} < \trh$). 
Curves of constant relic density will then follow a steeper
slope for the bremsstrahlung process than for the annihilation, and this corresponds to the result we obtained 
numerically in Fig.~\ref{fig:gen_compare}.
Note also the change of sign in the slope for $n=0$ when one reaches
the point $m_{\cal X}=\trh$. This comes from the fact that, for a given mass, the time needed to produce dark matter is reduced (from $\tmax$ to $m_{\cal X}$ rather than from $\tmax$ to $\trh$). In other words, to produce the same amount of dark matter, one needs larger $\tmax$, and as a consequence, larger $\trh$. The slope recovers the $m_{\cal X}\propto 1/{\trh}$ behavior characteristic of inflaton decay, when $R_{\phi \cal X}> R_{{\cal P} \cal X}$.

It is also interesting to note the change in behavior of the ratio $n_{{\cal X}}^{{\rm brem}}/n_{{\cal X}}^{{\rm ann}}$ as function of the dark matter mass.
Indeed, whereas the bremsstrahlung production depends only
on the reheating temperature (assuming we are well below the kinematic threshold), thermal production for $n<6$ depends strongly on the lower bound of the integration in Eq.~(\ref{eq:DM_prod_approx}), $T_c$. As long as $m_{\cal X} < \trh$, the lower bound is $\trh$ and the thermal production is independent of $m_{\cal X}$, as we can see in Fig.~\ref{fig:gen_compare} with $n=2$, $4$ and $6$, where this regime corresponds
to the horizontal red dashed lines. However, if $m_{\cal X}> \trh$, 
the thermal production is less efficient because it takes place over a shorter period of time, and for larger values of $m_{\cal X}$, the period is shorter. This explains why,
in the case of large dark matter masses, 
the bremsstrahlung production largely dominates the thermal one. 
To be more precise, For $T_{\rm max} \gtrsim m_{{\cal X}}\gtrsim \trh$,
the production from plasma annihilation is suppressed by a factor of $\trh^4/m_{\cal X}^4$ due to the production threshold at $T\simeq m_{\cal X}$, whilst for  bremsstrahlung production,
the only threshold is $2m_{{\cal X}}<m_{\phi}$. This suppression factor stems from the factor of $T_c^{6-n}$ in Eq.~(\ref{eq:DM_n_ratio}a).

This domination of bremsstrahlung production is 
 more pronounced for higher $n$ as can be seen in Fig.~\ref{fig:gen_compare} for  $n=4$ and $6$. For $m_\chi < \trh$ and $n\leq 6$ the ratio in Eq.~(\ref{eq:DM_n_ratio}) is proportional to 
$\left({m_\phi}/{\trh}\right)^n$. For $n=6$, bremsstrahlung production nearly dominates all of the parameter space with $\trh<m_\phi$.
For even larger values of $n>6$, 
${n_{{\cal X}}^{{\rm brem}}}/{n_{{\cal X}}^{{\rm ann}}}$ is independent of $m_\chi$ (so long as it is kinematically allowed)  and receives additional enhancement compared to $n\leq 6$ as is evident from Eq.~(\ref{eq:DM_n_ratio}c).
This ratio can easily reach $10^7$ for typical values of $\frac{\tmax}{\trh}=100$.
We conclude therefore that not only is bremsstrahlung production dominant for reheating temperatures $\trh \lesssim 10^{10}$ GeV in the case $n=2$ (corresponding to the exchange of a massive particle) but the dominance is even stronger for larger values of $n$. This is another important result of our work. We recall as well, that for $n=0$, bremsstrahlung dominates when $m_\chi < \trh$.

Thus, bremsstrahlung production dominates in a wide range of reheating temperatures.
Combining Eqs.~(\ref{Eq:brbrem}) 
and Eq.~(\ref{eq:brem_prod})
while writing the 
relic abundance as \cite{book}
\begin{equation}
\Omega_{\cal X}h^2 = 1.6\times 10^8 \, \frac{g_0}{g_{*}}\frac{n(\trh)}{\trh^3}\frac{m_{\cal X}}{1~{\rm GeV}}\,,
\label{oh2}
\end{equation}
with $g_0 = 3.91$, we obtain 
\begin{equation}
\frac{\Omega^{\rm brem}_{\cal X} h^2}{0.12} \simeq 
\tilde{c} \,
\frac{m_{\cal X}}{{2.3}\times 10^{8}\,{\rm GeV}}
\left( \frac{m_\phi}{10^{13}\,{\rm GeV}} \right)^3 
\frac{\trh}{10^{10} \, {\rm GeV} }
\left(\frac{10^{16} {\rm GeV} }{ \Lambda }\right)^4\, ,
\label{eq:relic_estimate0}
\end{equation} 
or
\begin{equation}
\frac{\Omega^{\rm brem}_{\cal X} h^2}{0.12} \simeq 
\tilde{c} \,
\frac{m_{\cal X}}{{7.0}\times 10^{10}\,{\rm GeV}}
\left( \frac{m_\phi}{10^{13}\,{\rm GeV}} \right)^3 
\frac{\trh}{10^{10} \, {\rm GeV} }
\left(\frac{10^{16} {\rm GeV} }{ M }\right)^4\, ,
\label{eq:relic_estimate}
\end{equation} 
 for $n=2$ where we have normalized to the observed cold dark matter density \cite{Planck:2018vyg}.
The threshold condition $m_{\cal X} < m_\phi \simeq 10^{13} \,{\rm GeV}$ sets $\Lambda\lesssim 4 \times 10^{17}$\,GeV for $\trh < 10^{12}$ GeV.
Therefore, to achieve the observed DM abundance, the parameter $\Lambda$ of the
interaction can lie between $m_\phi$ and
$4 \times 10^{17}$\,GeV. Note that this range is largely independent of the reheating temperature, since for a given $m_{\cal X}$, $\Lambda \propto \trh^\frac{1}{4}$. Within the viable parameter space according to this estimate, probably the most interesting  choice
for the interaction scale is the GUT scale, $\Lambda\simeq10^{16}\,$GeV.
In this case, the dark sector and the visible sector
is only connected by GUT scale interactions such as through heavy
gauge bosons of a unification group. This induces vector-type effective
interaction between the plasma particle and the DM. 
As one can see from Eq.~(\ref{eq:relic_estimate}), the relic density (for $n=2$) depends on four parameters, $m_{\cal \chi}, m_\phi, \trh$, and $M$. Lower dark matter masses can be obtained raising $\trh$ and lowering $M$ for fixed inflaton mass, $m_\phi=10^{13}$\,GeV. However if we raise $\trh$, the bremsstrahlung process will no longer dominate. Thus the lowest dark matter mass determined by bremsstrahlung is obtained for $\trh \simeq 10^{10}$ GeV and $M \gtrsim m_\phi$ giving $m_{\cal X}\gtrsim 70$\,MeV, for $\tilde{c}\sim 1$.

For a more general inflaton mass, the bremsstrahlung production may still dominate for sufficiently small $\trh$ and/or $\trh\ll m_{\cal X}$. However, the DM relic density poses a lower bound to $m_\phi$ because of the production threshold $m_\phi > 2m_{\cal X}$. Imposing $m_\phi \gg m_{\cal X}/2,\, \trh$, Eq.(\ref{eq:relic_estimate}) sets 
\begin{equation}
m_\phi \gtrsim 1.1\times 10^{12}\,{\rm GeV}
\left( \frac{M}{10^{16} \,{\rm GeV}} \right)^{4/5}
\end{equation}
for bremsstrahlung to dominate the production of the observed DM relic.

The estimates we have made up to now are mostly qualitative in that the model dependent factor $\tilde{c}$ is not guaranteed to be ${\cal O}(1)$.  
In the remainder of the paper, we perform numerical
calculations for the rate of DM bremsstrahlung and plasma annihilation
for dark matter produced through a gravity portal or a vector portal. We will see that the result
is in close agreement with the estimate of $\tilde{c}\sim{\cal O}(1)$
in this section.

\section{Gravitational bremsstrahlung
\label{sec:GR_production}}

\begin{figure}[t]
\centering
\includegraphics[width=0.9\textwidth]{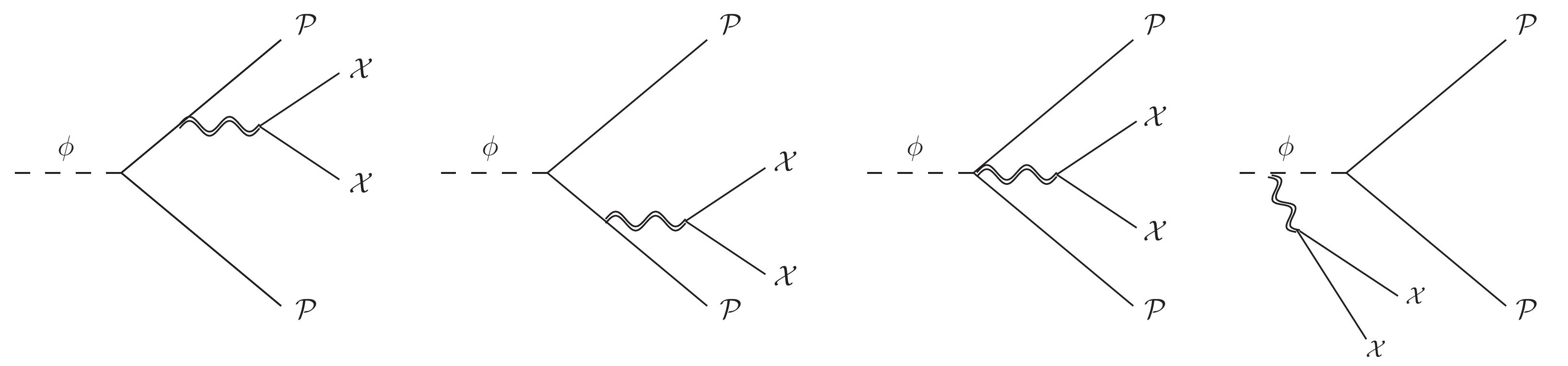} \caption{\label{fig:GR4-body}.The diagrams for gravitationally induced DM bremsstrahlung from inflaton decay.}
\end{figure}

A well motivated choice for the scale $\Lambda$ is the Planck scale. 
Indeed, DM will always interact with the inflaton or plasma particle through the exchange of a graviton,
$h_{\mu \nu}$. 
In this section, we evaluate the dark matter abundance
produced by gravitational bremsstrahlung that is unavoidable in inflationary cosmology. As the inflaton decays,
DM particles can be emitted from the decay products or from the inflaton
itself through a virtual graviton, leading to the 4-body final state,
\begin{equation}
\phi\rightarrow ss^{*}(\text{or}\,ff)+XX(\text{or}\,\bar{\chi}\chi)\,,
\end{equation}
where 
we denote by $X$ a real scalar, or $\chi$, a
Dirac fermion as the dark matter candidate.
The relevant diagrams are shown in Fig.~\ref{fig:GR4-body}. 
The analysis can be easily generalized to complex scalars or Majorana
fermions as DM by counting degrees of freedoms. We note that
contracting the two plasma particle lines in Fig.~\ref{fig:GR4-body} to close a loop 
generating a process $\phi \rightarrow {\cal P}_{\rm loop} \rightarrow h_{\mu \nu} \rightarrow {\cal X}~\cal{X} $ (depicted in Fig.~\ref{fig:DM_production}(c) replacing the effective coupling by a graviton)
gives zero since the spin-$2$ massless graviton cannot mix with the spin-$0$ inflaton. 
This is in contrast to the cases treated
in \cite{Kaneta:2019zgw}.
Therefore, the gravitational bremsstrahlung is the leading {\it radiative} production process of dark matter. We
will show that even when bremsstrahlung is the dominant gravitational production
mechanism in some parts of the parameter space, it cannot saturate
the observed dark matter abundance alone. 
In general, the dominant gravitational source of dark matter comes from the gravitational annihilation of inflaton \cite{Mambrini:2021zpp,Clery:2021bwz}.
In a sense, our result confirms the approximate 
result obtained in the previous section, since the
exchange of a graviton can be seen as
an effective model with $\Lambda=M_P > 10^{17}$ GeV.

To evaluate the bremsstrahlung rate, we use the Monte Carlo event generator {\tt MadGraph}\,\cite{Alwall:2011uj} to 
calculate the matrix element and simulate the final 4-body phase space. {\tt MadGraph} supports all of the Lorentz structure of vertices\,\cite{deAquino:2011ub} for a massless graviton\footnote{The massive graviton propagator does not reduce to the massless one in the zero mass limit because of the extra degrees of freedom.~\,\cite{vanDam:1970vg,Zakharov:1970cc}} that appear in Fig.~\ref{fig:GR4-body} which is crucial in the cancellation of collinear divergences.
We use
{\tt FeynRules}\,\cite{Alloul:2013bka} to generate the UFO model files for {\tt MadGraph}.

Some comments on the infrared behavior of the process is in order to justify
the validity of the numerical simulation. A massless mediator may lead to soft or collinear divergences like in QED or QCD which needs to be regulated with a cut in the particle momenta. Fortunately, gravity
has no collinear divergences \cite{Weinberg:1965nx} and
thus no subtle treatment is required. Indeed,
we have checked analytically that all of the soft and collinear divergences in Fig.~\ref{fig:GR4-body} cancel but only
after summing over all of the four contributing diagrams. This introduces subtleties in numerical evaluation with {\tt MadGraph}. By default, {\tt MadGraph}
handles the particle widths with Breit-Wigner propagators.%
\,\footnote{Many thanks to Olivier Mattelaer for pointing this out in the {\tt MadGraph5\_aMC@NLO}
	question and answer launchpad: \url{https://answers.launchpad.net/mg5amcnlo/+question/698109}}
The width in denominators of these propagators mix
contributions of different orders of the perturbative calculation and spoils the delicate cancellation of the IR divergences.
To maintain the cancellation at a fixed order, we set the widths of all particles to
zero when simulating with {\tt MadGraph}. The cancellation of large numbers from almost on-shell propagators between diagrams could pose another challenge to the numerical stability. To evade this issue, we applied a cut on the DM invariant mass, $(\sum p_{{\rm DM}})^{2}>10^{-8}m_{\phi}^{2}$,
to get rid of very low energy DM in the numerical simulation. Due
to the absence of soft divergences, the DM particles are rarely produced with soft or near collinear momenta. Thus, this cut guarantees the numerical
stability with only mild effects on the resultant inflaton partial width.~\footnote{Practically, such a cut is not even needed in a wide range of inflaton/DM
mass ratios after scaling down $M_{P}$ thanks to the numerical precision of
{\tt MadGraph}. This can be used as a stringent consistency
check of the implementation of the model.} The numerical result for the inflaton partial width due to the DM bremsstrahlung,
in the limit of $m_{{\cal X}}\ll m_{\phi}$ is
\begin{align}
\Gamma_{\phi,{\rm brem}}^{G} & ={\cal C}_{{\cal PX}}^{G}\lambda_i^{2}\frac{m_{\phi}^{5}}{{M}_{P}^{4}}\,.~\label{eq:Rate_Brem}
\end{align}
where the coupleings $\lambda_{i=s,f}$ are given in Eq.~(\ref{Eq:lambdai}).
We can now evaluate the coefficients ${\tilde c}$ for $M = M_P$ and $n=2$,
\begin{equation}
{\tilde c} = 2^{16} \pi^5 {\cal C}_{{\cal PX}}^{G}
\,,
\end{equation}
and the numerical factor ${\cal C}_{{\cal PX}}^{G}$ is obtained
by simulation,
\begin{subequations}
\begin{align}
{\cal C}_{sX}^{G} & \simeq2.4\times10^{-9}\,,\\
{\cal C}_{s\chi}^{G} & \simeq4.1\times10^{-9}\,,\\
{\cal C}_{fX}^{G} & \simeq2.6\times10^{-9}\,,\\
{\cal C}_{f\chi}^{G} & \simeq4.3\times10^{-9}\,,
\end{align}    
\end{subequations}
for $\phi\rightarrow ss^{*}XX$, $ss^{*}\bar{\chi}\chi$, $ffXX$,
and $ff\bar{\chi}\chi$, respectively. 
From Eq.~\eqref{eq:relic_estimate}, we see that the gravitational bremsstrahlung process
alone cannot produce enough DM particles
to saturate the observed
relic abundance when $M = M_P$ and $m_{\cal X},\trh < m_\phi$. 

The dark matter particle can also be produced gravitationally
by the annihilation of the inflaton during its oscillationary epoch or by the
annihilation of a pair of plasma particle during or after reheating.
These two scenarios have been considered widely in previous works such
as \cite{Tang:2017hvq,Ema:2015dka,Ema:2018ucl,Ema:2019yrd,Mambrini:2021zpp,Clery:2021bwz}.
The production from inflaton annihilation was found to be dominant
whenever the gravitational production saturates the observed abundance\footnote{see also \cite{Garcia:2022vwm,Kaneta:2022gug} for an alternative treatment of the gravitational production.}.
The gravitational annihilation of the inflaton may also leave a significant impact on the maximal or the reheating temperature of the thermal plasma\,\cite{Clery:2021bwz,Haque:2022kez,Co:2022bgh,Clery:2022wib}.
We briefly summarize the result for inflaton annihilation \cite{Mambrini:2021zpp,Clery:2021bwz}
as follows. For inflaton annihilation  $\phi\phi\rightarrow XX$ or
$\bar{\chi}\chi$ during oscillations about a quadratic potential, the DM production rate that enters the Boltzmann
equation Eq.~(\ref{eq:DM_prod}) is
\begin{equation}
R_{\phi\phi}^{G}=\frac{\rho_{\phi}^{2}}{128\pi{M}_{P}^{4}}\beta_{\phi\phi}\,,
\end{equation}
where $\beta_{\phi\phi}=1$ for $\phi\phi\rightarrow XX$ and $\beta_{\phi\phi}=m_{\chi}^{2}/m_{\phi}^{2}$
for $\phi\phi\rightarrow\bar{\chi}\chi$, assuming $m_{\chi,X} \ll m_\phi$. All possible DM degrees
of freedom have been summed in this inclusive production rate. The
production of fermionic DM is suppressed by its mass compared to the
case of scalar DM since the DM particle, anti-particle pair must
have the same helicity or the opposite chiralities. In contrast, the
production rate from bremsstrahlung does not suffer from this suppression because the pair of plasma particles ($ss^{*}$
or $ff$ ) in the decay product carry angular momentum so fermionic
DM does not have to be in the same helicity state. The produced DM density
at $\trh$ is \cite{Mambrini:2021zpp,Clery:2021bwz}
\begin{equation}
n_{{\cal X}}^{\phi\phi}(\trh)\simeq\frac{\sqrt{3}\rho_{\rm RH}^{\frac{3}{2}}}{192\pi{M}_{P}^{3}}\left(\frac{\rho_{{\rm end}}}{\rho_{\rm RH}}\right)^{\frac{1}{2}}\beta_{\phi\phi}\,,
\label{Eq:nphigrav}
\end{equation}
where $\rho_{{\rm end}}$ is the energy density stored in the inflaton
field at the end of the slow roll inflation. At the start of inflaton
oscillations, the high energy density of the inflaton condensate significantly
enhances the production rate of dark matter. Using Eq.~(\ref{oh2}) and Eq.~(\ref{Eq:nphigrav}), we obtain~\cite{Mambrini:2021zpp,Clery:2021bwz}
\begin{equation}
\frac{\Omega_{\chi}^{\phi \phi} h^2}{0.12}\simeq
\left(\frac{10^{13}~ {\rm GeV}}{m_\phi}\right)^2
\left(\frac{\trh}{10^{10} ~{\rm GeV}}\right)
\left(\frac{\rho_{\rm end}}{10^{64} ~{\rm GeV}^4}\right)^{\frac{1}{2}}
\left(\frac{m_\chi}{6.4\times 10^{10}~ {\rm GeV}}\right)^3
\label{Eq:omegaphihalf}
\end{equation}
for
the relic abundance
of a fermionic dark matter candidate, $\chi$. 
The range of DM masses for which the bremsstrahlung process dominates the production of dark matter relative to inflaton scattering is found by
comparing Eqs.~(\ref{Eq:omegaphihalf}) and (\ref{eq:relic_estimate}). 
For ${\tilde c} \simeq 0.08$,  this leads to
\begin{equation}
m_\chi \lesssim 0.3 ~\left(\frac{M_P}{M}\right)^2
\left(\frac{m_\phi}{10^{13}~{\rm GeV}}\right)^{5/2}
\left(\frac{10^{16} ~{\rm GeV}}{\rho_{\rm end}^{\frac{1}{4}}}\right)~{\rm PeV}\,,
\end{equation}
However, for these masses, the production is much too low 
to match the CMB-inferred relic density \cite{Planck:2018vyg}, as we can see from 
Eq.~(\ref{eq:relic_estimate}). The production of scalars, $X$, is even less efficient since the gravitational inflaton
scattering does not depend on the mass of the dark matter, and hence always dominates over the bremsstrahlung process. We can then safely neglect
the gravitational bremsstrahlung process for the rest of our work.

\section{Vectorial bremsstrahlung}
\label{sec:Vector_Portal}

In GUT-inspired models, there are typically gauge bosons with mass $\gtrsim 10^{15}$ GeV, i.e., much above the mass of the inflaton or $\tmax$.
This type of heavy mediator also appears in see-saw
constructions.
In this section, we proceed to discuss DM bremsstrahlung during inflaton
decay in the general cases where the DM ${\cal X}$ and the plasma
particle ${\cal P}$ interact through a heavy vector mediator. 
The dark matter candidate can be
a \emph{complex} scalar ${\bf X}$ or a Dirac spinor $\chi$. We will stay within the limit of $m_{{\rm {\cal X}}}\ll m_{\phi}$,
i.e., the decay products of the inflaton are effectively massless
in the amplitude. 

We consider the following dimension-six vector portal effective interactions suppressed by a cut-off scale $M\gg m_\phi$, 
\begin{subequations}
\begin{align}
{\cal L}_{s{\bf X}}^{V} & =
\frac{g_{s {\bf X} }^2}{M^{2}}
\left(s^{*}\partial^{\mu}s-s\partial^{\mu}s^{*}\right)\left({\bf X}^{*}\partial_{\mu}{\bf X}-{\bf X}\partial_{\mu}{\bf X}^{*}\right)\,,\\
{\cal L}_{s\chi}^{V} & =\frac{g_{s\chi}^2}{M^{2}}{\rm i}\left(s^{*}\partial_{\mu}s-s\partial_{\mu}s^{*}\right)\bar{\chi}\gamma_{\mu}\chi\,,\\
{\cal L}_{f {\bf X} }^{V} & =
\frac{g_{f {\bf X}}^2}{M^{2}}
{\rm i}\bar{f}\gamma_{\mu}f\left({\bf X}^{*}\partial_{\mu}{\bf X}-{\bf X}\partial_{\mu}{\bf X}^{*}\right)\,,\\
{\cal L}_{f\chi}^{V} & =\frac{g_{f\chi}^2}{M^{2}}\bar{f}\gamma_{\mu}f\bar{\chi}\gamma_{\mu}\chi\,.
\end{align}
\label{eq:vec_int}
\end{subequations}
The fermionic bilinears with $\gamma_{5}$ are equivalent to the ones without it in the massless limit since the phases of chiral spinors can be redefined independently. 
These interactions can be obtained by integrating out a heavy gauge
boson that interact with both the DM particle and the plasma particle as we will demonstrate briefly at the end of this section.
In such a scenario, the real scalar inflaton condensate
cannot annihilate efficiently through the vector portal because it requires the presence of
another ``partner scalar'' field to form a complex scalar to interact
with a gauge boson at the tree level.
Although the inflaton is often the real part of a complex field, the imaginary part is far less
abundant during inflaton oscillations. Thus, the annihilation of the inflaton
to DM can only occur through one-loop diagrams with two virtual gauge
bosons and one virtual ``partner scalar'', and is thus suppressed
by at least an additional factor of $m_{\phi}^{4}/M^{4}$. Moreover,
the radiative decay of the inflaton to dark matter $\phi\rightarrow{\bf X^{*}X}$
or $\phi\rightarrow\bar{\chi}\chi$ induced by a plasma particle loop as in Fig.~\ref{fig:DM_production}(c)
vanishes for vector type effective interactions because a massive scalar cannot mix with a vector mediator. Technically, since the plasma particle loop only has one Lorentz index, it must be proportional
to the inflaton 4-momentum $p_{\phi}^{\mu}$ and vanishes after contracting
with the current of the final state DM particles.  
Thus, as in the case of gravitational portal, bremsstrahlung is the leading {\it radiative} production mechanism.

Following the same procedure as in the previous section, the DM bremsstrahlung rate of $\phi\rightarrow{\cal PPXX}$ induced
by the above interactions for $M \gg m_{\phi}\gg m_{s},m_{f}$
is parameterized  as,
\begin{equation}
\Gamma_{\phi\rightarrow{\cal PPXX}}^{V}={\cal C}_{{\cal PX}}^{V}g_{{\cal PX}}^{4}\lambda_i^{2}\frac{m_{\phi}^{5}}{M^{4}}\,,
\end{equation}
with ${\cal C}_{{\cal PX}}^{V}$ a numerical factor and $g_{{\cal PX}}$
the dimensionless coefficient of the effective operators as defined
in Eq.~\eqref{eq:vec_int}.
In this case,
\begin{equation}
{\tilde c} = 2^{16} \pi^5{\cal C}_{{\cal PX}}^{V}g_{{\cal PX}}^{4} \, .
\end{equation}
As we did for the gravitational coupling,
we use {\tt MadGraph} for numerical simulation of the 4-body decay. We obtained
\begin{subequations}
\begin{align}
{\cal C}_{s{\bf X}}^{V} & =1.0\times10^{-8}\,,\\
{\cal C}_{s\chi}^{V} & =4.1\times10^{-8}\,,\\
{\cal C}_{f{\bf X}}^{V} & =6.9\times10^{-9}\,,\\
{\cal C}_{f\chi}^{V} & =2.8\times10^{-8}\,,
\end{align}
\end{subequations}
for $\phi\rightarrow ss^{*}{\bf X}{\bf X}^{*}$, $\phi\rightarrow ss^{*}\chi\bar{\chi}$,
$\phi\rightarrow f\bar{f}{\bf X}{\bf X}^{*}$ and $\phi\rightarrow f\bar{f}\chi\bar{\chi}$,
respectively. 

On the other hand, the rate of DM production by annihilation 
of particles from the plasma ${\cal P}$
is evaluated as\,\cite{Clery:2021bwz} 
\begin{equation}
R^{\cal P}_{\rm ann} = \sum_{i_1,i_2} \frac{1}{1024\pi^6}
\int {\rm d}E_1 {\rm d}E_2 E_1 E_2 f_{i_1}(E_1) f_{i_2}(E_2) {\rm d}\cos\theta_{12} {\rm d}\Omega_{13}
\left| {\cal M}^{i_1 i_2} \right|^2 \,,
\end{equation}
where $i_1$ and $i_2$ run over all annihilating particle species. The angles and the energies are related to the Mandelstam variables by $t=s(\cos\theta_{13}-1)/2$ and $s=2E_1 E_2 (1-\cos\theta_{12})$. For the vector portal interactions parametrized by Eqs.~\eqref{eq:vec_int}, the production
rates are given by
\begin{subequations}
	\begin{align}
       R_{\rm ann}^{s{\bf X}} &= 
       \frac{g_{s {\bf X}}^4 \pi^3 T^8 }{8100M^4}\,,\\
       R_{\rm ann}^{s{\chi}} &= 
       \frac{g_{s {\chi}}^4 \pi^3 T^8 }{2025M^4}\,,\\
       R_{\rm ann}^{f{\bf X}} &= 
       \frac{49 g_{f {\bf X}}^4 \pi^3 T^8 }{259200M^4}\,,\\
       R_{\rm ann}^{f{\chi}} &= 
       \frac{49 g_{f {\chi}}^4 \pi^3 T^8 }{64800M^4}\,.
	\end{align}
\end{subequations}

\begin{figure}[t]
	\centering
	\includegraphics[width=0.37\textwidth]{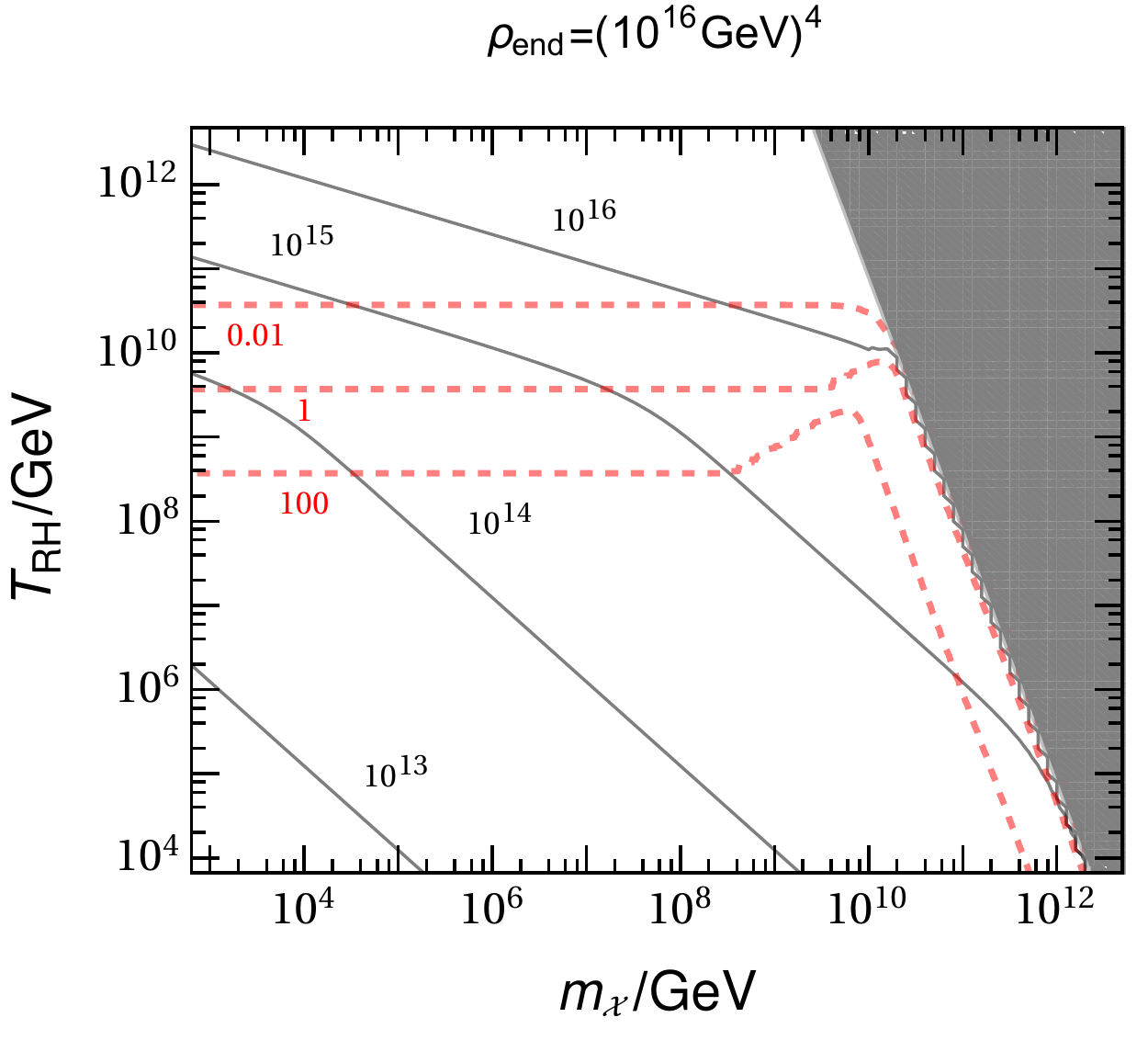}%
	\hspace*{5mm}
	\includegraphics[width=0.37\textwidth]{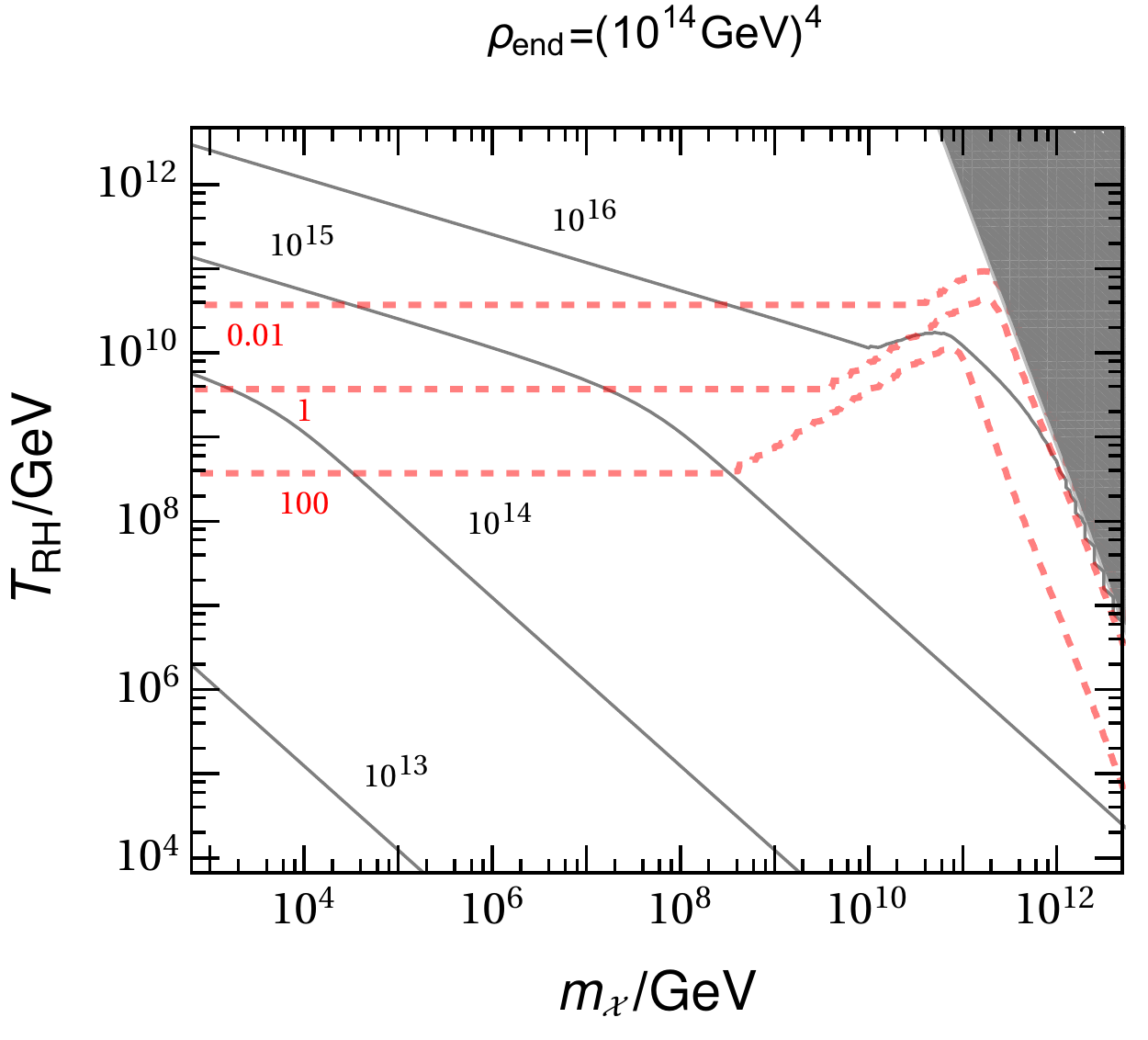}%
	\\
	\includegraphics[width=0.37\textwidth]{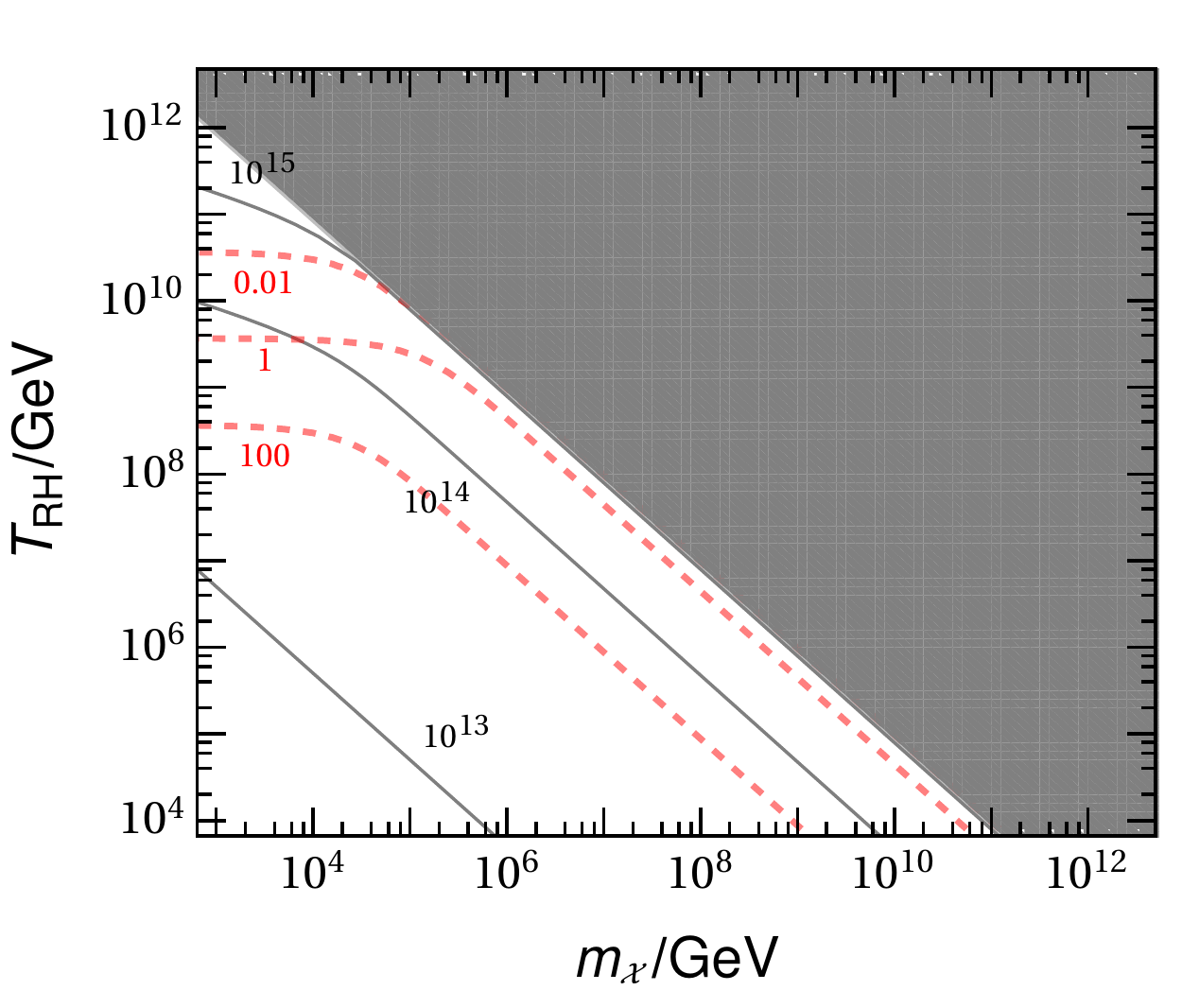}%
	\hspace*{5mm}
	\includegraphics[width=0.37\textwidth]{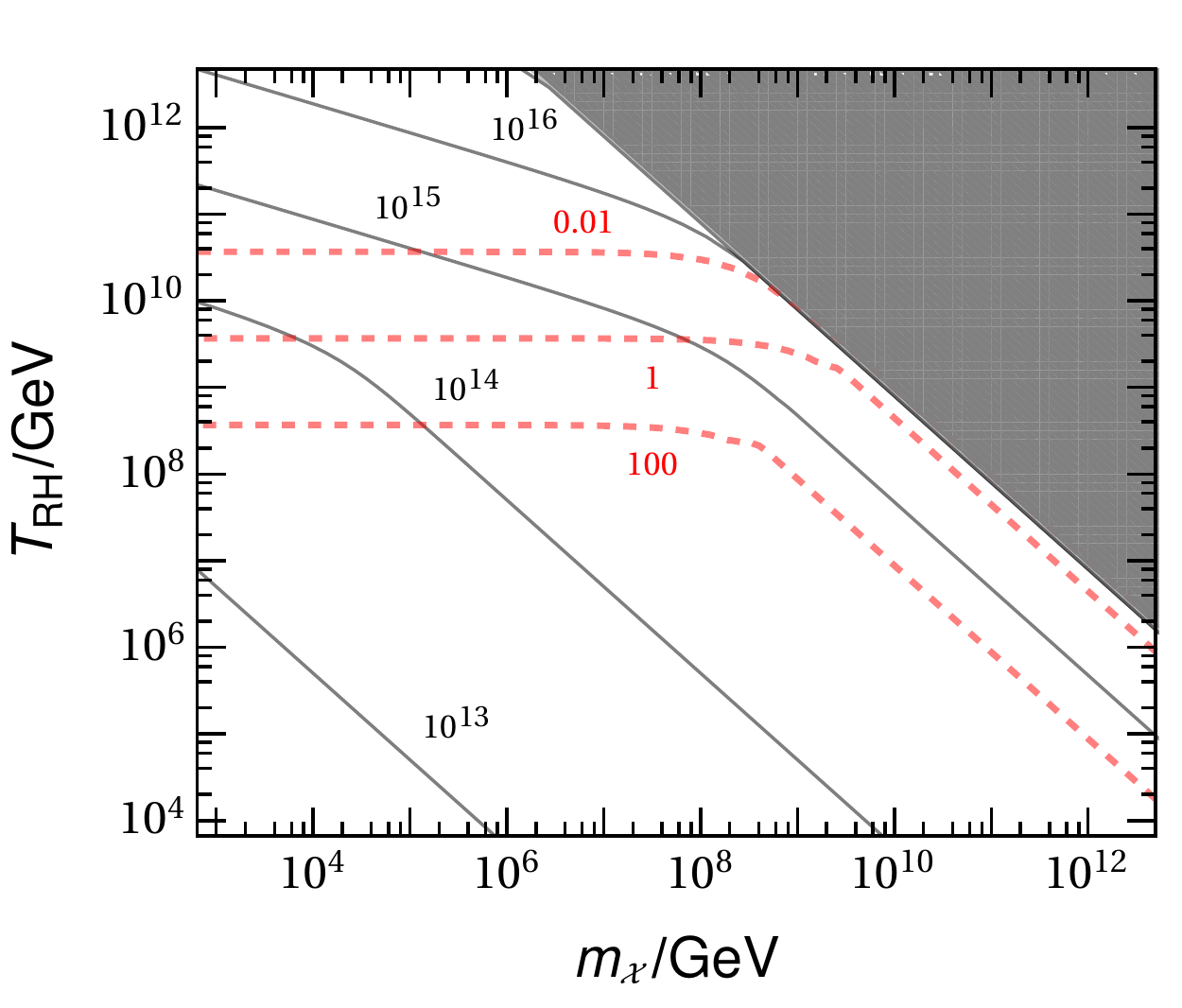}%
	\caption{\label{fig:vportal_compare}Parameter space that realizes the observed
		dark matter relic density for dimension-6 operators with $m_{\phi}=10^{13}$GeV.
		The upper and lower panels are for ${\cal L}_{f\chi}^{V}$ and ${\cal L}_{f{\bf X}}^{V}$,
		respectively. The left panels and right panels corresponds to $\rho_{{\rm end}}=(10^{16}{\rm GeV})^{4}$
		and $(10^{14}{\rm GeV})^{4}$. The black contours show the required
		effective interaction scale, $M/g_{f{\cal X}}$ in GeV. The red line presents the
		relative importance $n_{{\cal X}}^{{\rm brem}}/(n_{{\cal X}}^{{\rm other}})$
		of bremsstrahlung vs other production mechanisms combined for the
		choice of $M/g_{f{\cal X}}$ that achieves a realistic DM relic. The gray shaded
		region is where inflaton annihilation through gravity overproduces
		the DM. }
\end{figure}

We are now able to compare the three modes of dark matter production : the gravitational production through inflaton scattering, the bremsstrahlung production, and the scattering from particles in the plasma.
In Fig.~\ref{fig:vportal_compare}, we present the parameter space
for vector-type interactions that achieves the observed dark matter
relic density considering all production modes. 
The upper and lower
panels are for ${\cal L}_{f\chi}^{V}$ and ${\cal L}_{f{\bf X}}^{V}$,
respectively. The left panels and the right panels corresponds to $\rho_{{\rm end}}=(10^{16}{\rm GeV})^{4}$
and $(10^{14}{\rm GeV})^{4}$. In all cases, we assumed $m_{\phi}=10^{13}$ GeV
and $g_{*} = 100$. The black contours show the
required effective interaction scale $M/|g_{f{\bf X}}|$ or $M/|g_{f\chi}|$.
The gray shaded
region is where inflaton annihilation through gravity overproduces
the DM. 
As this contribution can not be removed, the shaded region is excluded. 
Along the boundary of the shaded region, the relic density produced by inflaton annihilations mediated by gravity matches the observed cold dark matter density without any additional contributions. 
The red lines are the contours when
the relative importance 
of bremsstrahlung vs other production mechanisms combined 
$n_{{\cal X}}^{{\rm brem}}/(n_{{\cal X}}^{{\rm other}})=0.01$, $1$, and $100$
for the choice of $M/g_{f{\cal X}}$ that attains a realistic DM relic density equal to that inferred from CMB observations \cite{Planck:2018vyg}. The red lines may have sharp turns at $m_{\cal X} = \trh$ due to the suppression of plasma annihilation at $m_{\cal X} > \trh$. 
As we have seen previously, the DM bremsstrahlung dominates at smaller
reheating temperatures of $\trh\lesssim10^{10}$GeV,
whereas to achieve the
observed DM relic density, the scale $M$ of the effective interaction
can lie in a wide range above $m_{\phi}$ up to $M\sim10^{16}$GeV, in accordance with our estimate in Section~\ref{sec:general consideration}. 
As expected, the allowed parameter space is larger
for lower values of $\rho_{\rm end}$ due to the smaller amount of dark matter produced gravitationally.

As a prove of concept, we outline a construction of a gauged $U(1)_{B-L}$ extension
of the Standard Model that contains all of the ingredients discussed in this section. The model can be realized as an intermediate subgroup
of the $SO(10)$ grand unification\,\cite{Fritzsch:1974nn,Georgi:1974my}. In this model, the $U(1)_{B-L}$
is spontaneously broken by the vacuum expectation value (vev) $v_s$ of a heavy
Higgs field $S$ with $(B-L)_{S}=2$, which 
can be the Majoron \cite{Rothstein:1992rh,Dudas:2014bca,Dudas:2020sbq}. The right-handed
neutrinos $N_{i}$ have $(B-L)_{N}=-1$ and obtain a large majorana
mass through its Yukawa coupling to $S$ as in the seesaw mechanism\,\cite{Yanagida:1980xy,Minkowski:1977sc,GellMann:1980vs,Mohapatra:1979ia,Schechter:1980gr,Schechter:1981cv}. In addition, we introduce a Dirac spinor $\chi$ with $(B-L)_{N}=-1$
as the dark matter and impose a $\mathbb{Z}{}_{2}$ symmetry for its stability.~\footnote{The $\mathbb{Z}{}_{2}$ symmetry can be realized as a remnant discrete
symmetry of the broken $SO(10)$ unification group\,\cite{Krauss:1988zc,Ibanez:1991hv,Ibanez:1991pr,Martin:1992mq}.} The inflaton is taken to be a real singlet $\phi$.
The interaction terms relevant to the discussion are
\begin{equation}
{\cal L}\supset\bar{\chi}{\rm i}D_{\mu}\gamma^{\mu}\chi+{\rm i}\bar{N}D_{\mu}\gamma^{\mu}N+\frac{y_{N}}{2}N^{T}CNS+\mu_s\left|S\right|^{2}\phi\,,\label{eq:vportal_model}
\end{equation}
where $D_{\mu}$ is the gauge covariant derivative and we have suppressed
the generation index of $N_{i}$. Expanding $S=v_s+s$, the $U(1)_{B-L}$
gauge boson $V$ obtains a mass $m_{V}=2\sqrt{2}g_{B-L}v_s$. For $m_{V}$
and $m_{s}$ much heavier than $m_{\phi}$, we can integrate out the
heavy gauge boson and the Higgs mode. The effective interactions become,
\begin{equation}
{\cal L}_{{\rm eff}}\supset\frac{g_{B-L}^{2}}{m_{V}^{2}}\bar{\chi}\gamma^{\mu}\chi\bar{N}\gamma_{\mu}N+\frac{m_{N}\mu_s}{m_{s}^{2}}N^{T}CN\phi\,,
\end{equation}
where $m_{N}\equiv y_{N}v_s$ is the Majorana neutrino mass. Here, we have
obtained the Yukawa operator responsible for reheating and the dimension-6
operator for dark matter production. Comparing with Eq.~\eqref{eq:vec_int}
and Eq.~\eqref{eq:L_reheating}, we can identify
\begin{subequations}
\begin{align}
\lambda_s & =2\frac{m_{N}\mu_s}{m_{s}^{2}}\,,\\
\frac{g_{f\chi}^2}{M^{2}} 
& = \frac{g_{B-L}^{2}}{m_{V}^{2}}\,.
\end{align}
\label{eq:vportal_eff}
\end{subequations}
In this case, we have checked that if one considers a Standard Mode-like Higgs mechanism, with $\mu_s \simeq m_s$,
for right handed neutrino masses below $\lesssim 10^{-4} m_s$
(leading to $\trh \lesssim 10^{10}$ GeV),
bremsstrahlung production radiated from a neutrino leg dominates all other type of dark matter production.
Note that in Eq.~\eqref{eq:vportal_model}, we only present terms
that respect the chiral symmetry of $\chi$. It is technically natural
to set the chiral symmetry breaking parameters $m_{\chi}$ and the coupling of $\bar{\chi}\chi\phi$ 
to small values to suppress the direct decay of $\phi$ to a DM pair. 
In this case, the dominant DM production process is either the plasma annihilation, the DM bremsstrahlung or the gravitationally annihilation of the inflaton, as we have presented in Fig.~\ref{fig:vportal_compare}.

\section{Conclusion and Discussion
\label{sec:conclusion}}

In this paper, we have computed the DM density produced by bremsstrahlung off an inflaton decay product.
We evaluated the rate of such production in general scenarios of DM - plasma particle interaction, including gravity and vector portal dimension-6 effective operators. The DM bremsstrahlung dominates when the reheating temperature is low so that the plasma annihilation to DM becomes less efficient. This significantly reduces the required dark matter mass to achieve the observed DM abundance for low reheating temperatures. 

We stress that such dark bremsstrahlung process generally occurs in the UV freeze-in DM scenario widely considered in the literature. It requires only two ingredients: 1) the decay of the inflaton to particles in the thermal plasma which is a general and natural assumption in inflationary cosmology and necessary for reheating; and 2) a feeble interaction between the DM and the plasma particle mediated by a heavy mediator, commonly found in unification theories and UV freeze-in DM models. Therefore, this effect needs to be considered in almost every UV freeze-in DM scenarios.
Besides the dark matter abundance, the dark bremsstrahlung can also produce very energetic dark radiation that could leave imprints in cosmological observations\cite{Ballesteros:2020adh}.

Finally, we note that the DM production can also be enhanced by the thermalization process\,\cite{Harigaya:2013vwa,Harigaya:2014waa,Garcia:2018wtq,Harigaya:2019tzu}. A full treatment of the thermalization process relies on the detail of the thermal plasma is beyond the scope of this paper. We include a qualitative comparison between the DM bremsstrahlung and the thermalization production to the appendix and conclude that while thermalization may compete with the production from thermal plasma, the DM bremsstrahlung always dominates the regime of low reheating temperature.

\section*{Acknowledgements}
We would like to thank S.~Clery, Y.~Xu, H.~He and N.~Bernal for useful discussions. This project has received support from the European Union's Horizon 2020 research and innovation programme under the Marie Skodowska-Curie Grant Agreement No 860881-HIDDeN and the IN2P3 Master Projet UCMN. The work of K.A.O.~was supported in part by DOE grant DE-SC0011842 at the University of Minnesota. J.Z.~was supported in part by the NSF of China\,(No.11675086 and 11835005).

\appendix
\section{Bremsstrahlung vs. Thermalization}

In the thermalization process, the decay products ${\cal P}$ of the inflaton condensate are still
in a very energetic state and may significantly increase the annihilation
cross section that scales as $E^{n}/\Lambda^{n+2}$. In this appendix, we briefly summarize the result for this process from the literature and compare it to DM bremsstrahlung. Ref. \cite{Garcia:2018wtq}
considered the annihilation of two energetic(hard) particles from the inflaton decay into a pair
of DM before the thermalization time scale $t_{{\rm th}}{\rm \approx\alpha^{-16/5}}m_{\phi}^{4/5}\Gamma_{\phi}^{-3/5}{M}_{P}^{-6/5}$.
The DM density generated by such a process by the end of reheating
can be recast as
\begin{equation}
n_{{\rm th},0}=c_{{\rm th}}\frac{\trh^{34/5}m_{\phi}^{n-6/5}}{{M}_{P}^{3/5}\Lambda^{n+2}}\,.
\end{equation}
For a rough order of magnitude estimate, the numerical factor $c_{{\rm th}}\lesssim{\cal O}(10^{9-n})$
for $g_{*}\approx100$. The relative importance between the DM bremsstrahlung and this pre-thermalization production scales as $m_{\phi}^{11/5}{M}_{P}^{3/5}\trh^{-14/5}$.
Thus, the DM bremsstrahlung always dominates at reheating temperature below $m_\phi$.

Throughout the reheating
process, the hard(energetic) decay products from inflaton decay can also annihilate
with particles in the plasma into a pair of DM. Ref. \cite{Harigaya:2019tzu}
found that for $m_{\cal X}^2\lesssim \trh m_\phi$, the relative importance of such a hard-plasma annihilation process
to the pre-thermalization hard-hard annihilation is
\begin{equation}
\frac{n_{\rm th}^\prime}{n_{{\rm th},0}} \sim\begin{cases}
\left(\frac{T_{{\rm th}}}{\trh}\right)^{3/2}
\left( \frac{m_\phi}{\trh} \right)^{1/2}
\,, & \text{for } n = 0\,,\\
\left(\frac{T_{{\rm th}}}{\trh}\right)^{3/2}
\left( \frac{m_\phi}{\trh} \right)^{(2-n)/2}
\,, & \text{for } 0 < n\leq 5\,,\\
\left( \frac{T_{\rm th}}{m_\phi} \right)^{(n-2)/2}\,, & \text{for }n>5\,,
\end{cases}
\label{eq:hard-plasma_prod}
\end{equation}
where
\begin{equation}
T_{{\rm th}}\equiv\alpha_{0}^{4/5}\left(\frac{\Gamma_{\phi}{M}_{P}^{2}}{m_{\phi}^{3}}\right)^{2/5}m_{\phi}
\end{equation}
is the thermalization temperature and $\alpha_{0}\sim{\cal O}(10^{-2})$
the fine structure constant of the interaction that thermalizes the inflaton decay product. 
Note that for $m_{\cal X}\gtrsim \trh$, 
the production is further suppressed by a factor of $(\trh m_\phi/m_{\cal X}^2)^{3/2}$ because of the production threshold. We stress that the discussion of thermalization here is an order of magnitude estimation at best.

\begin{figure}[t]
	\centering
	\includegraphics[width=0.37\textwidth]{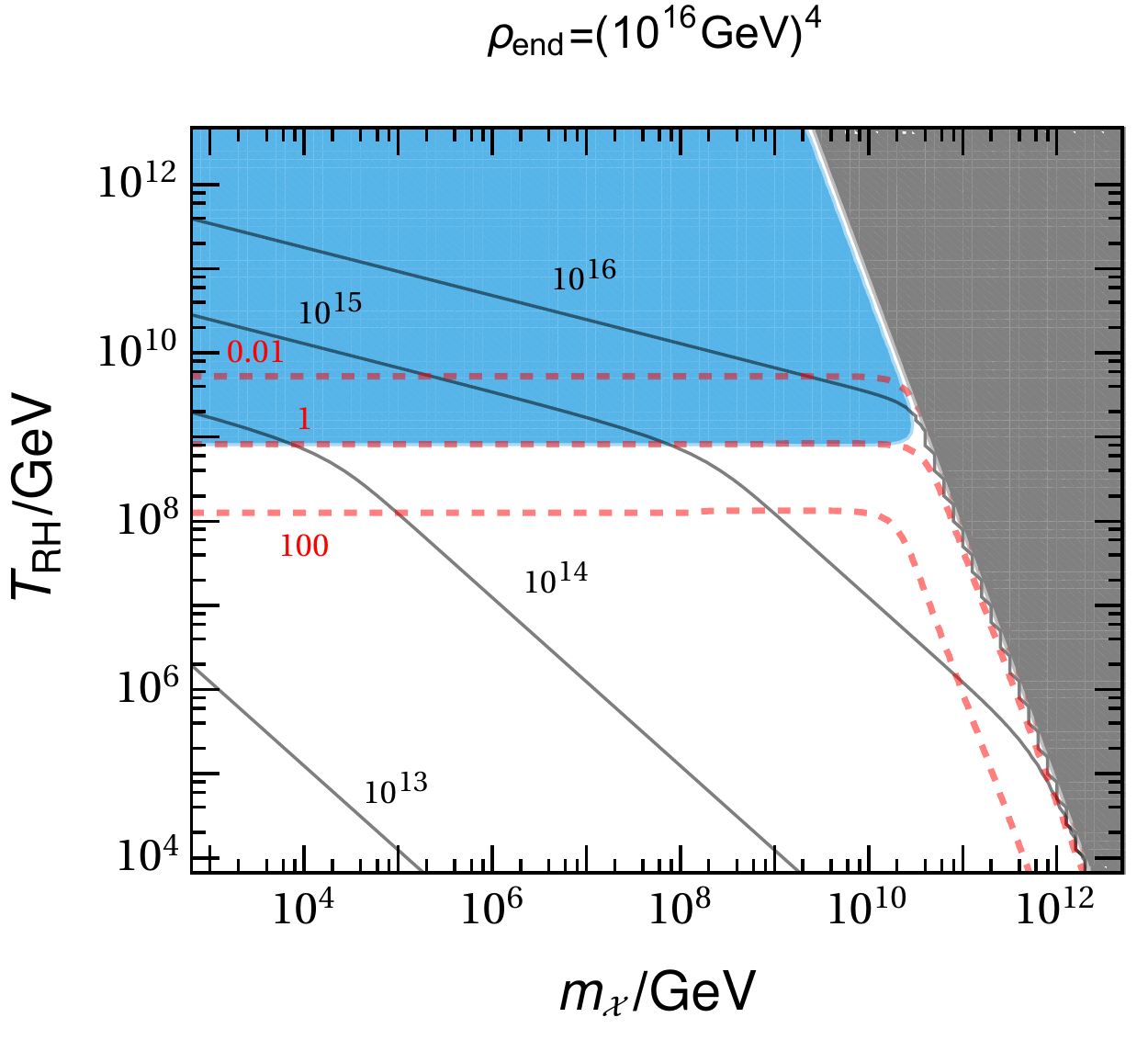}%
	\hspace*{5mm}
	\includegraphics[width=0.37\textwidth]{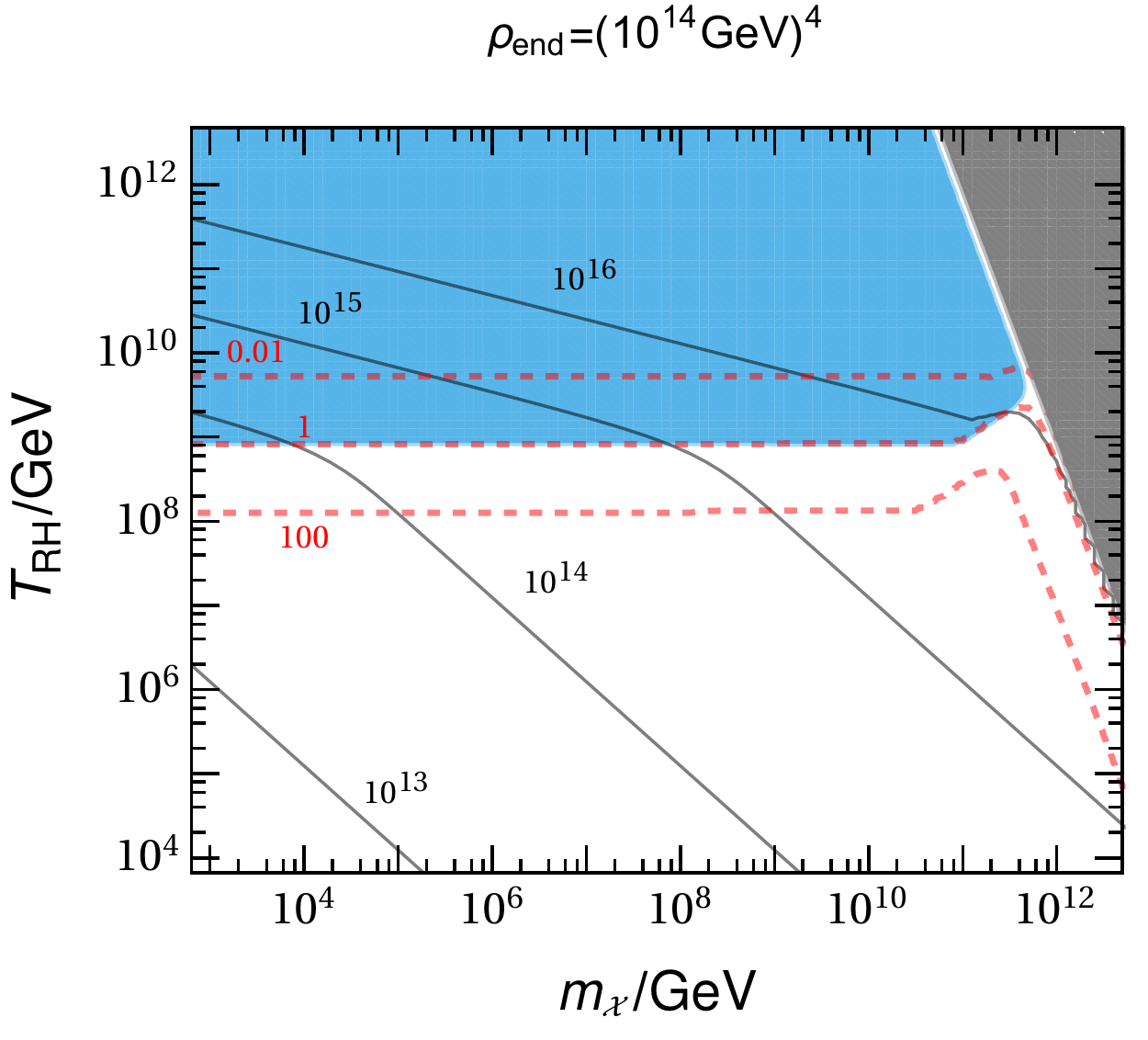}%
	\\
	\includegraphics[width=0.37\textwidth]{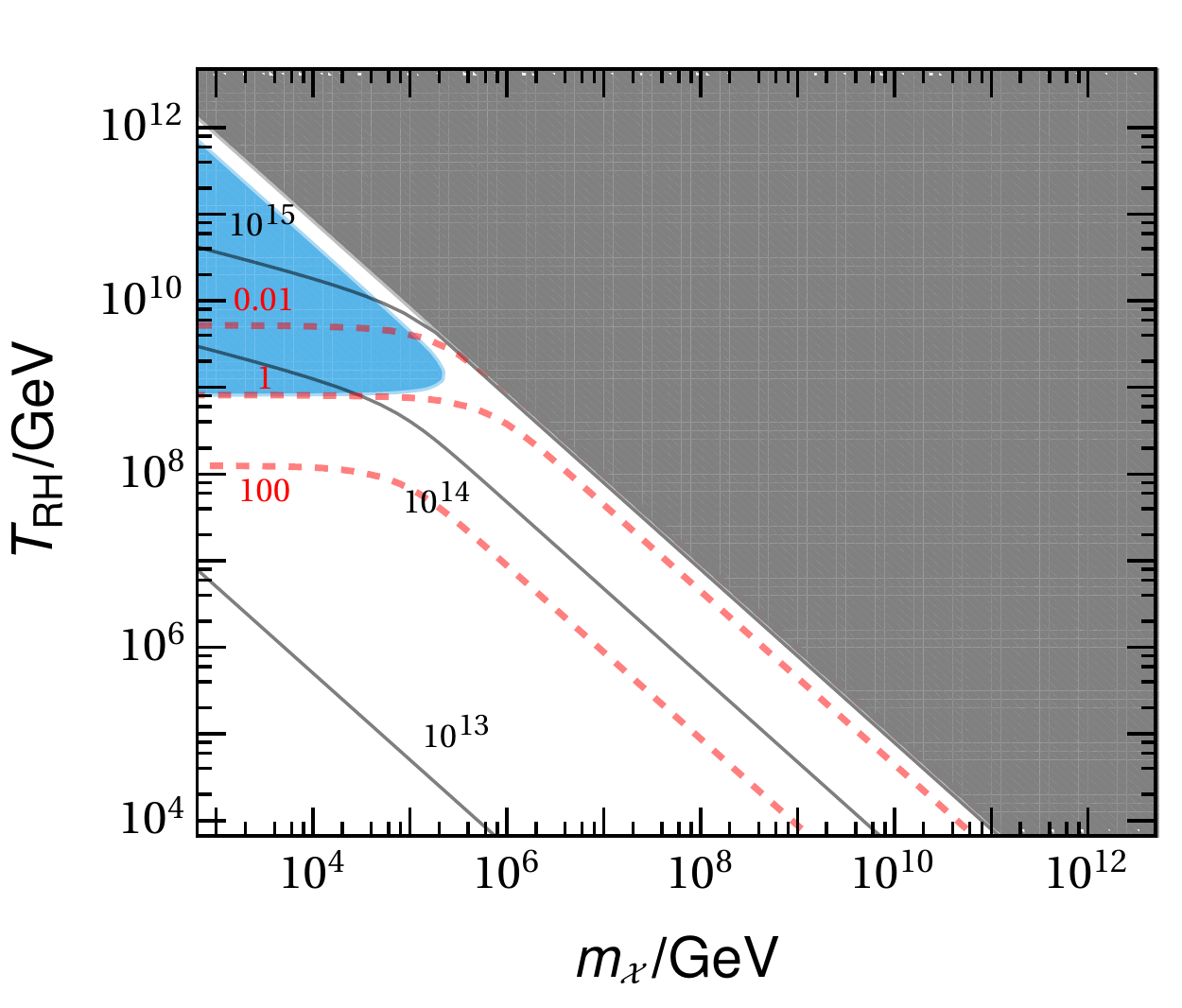}%
	\hspace*{5mm}
	\includegraphics[width=0.37\textwidth]{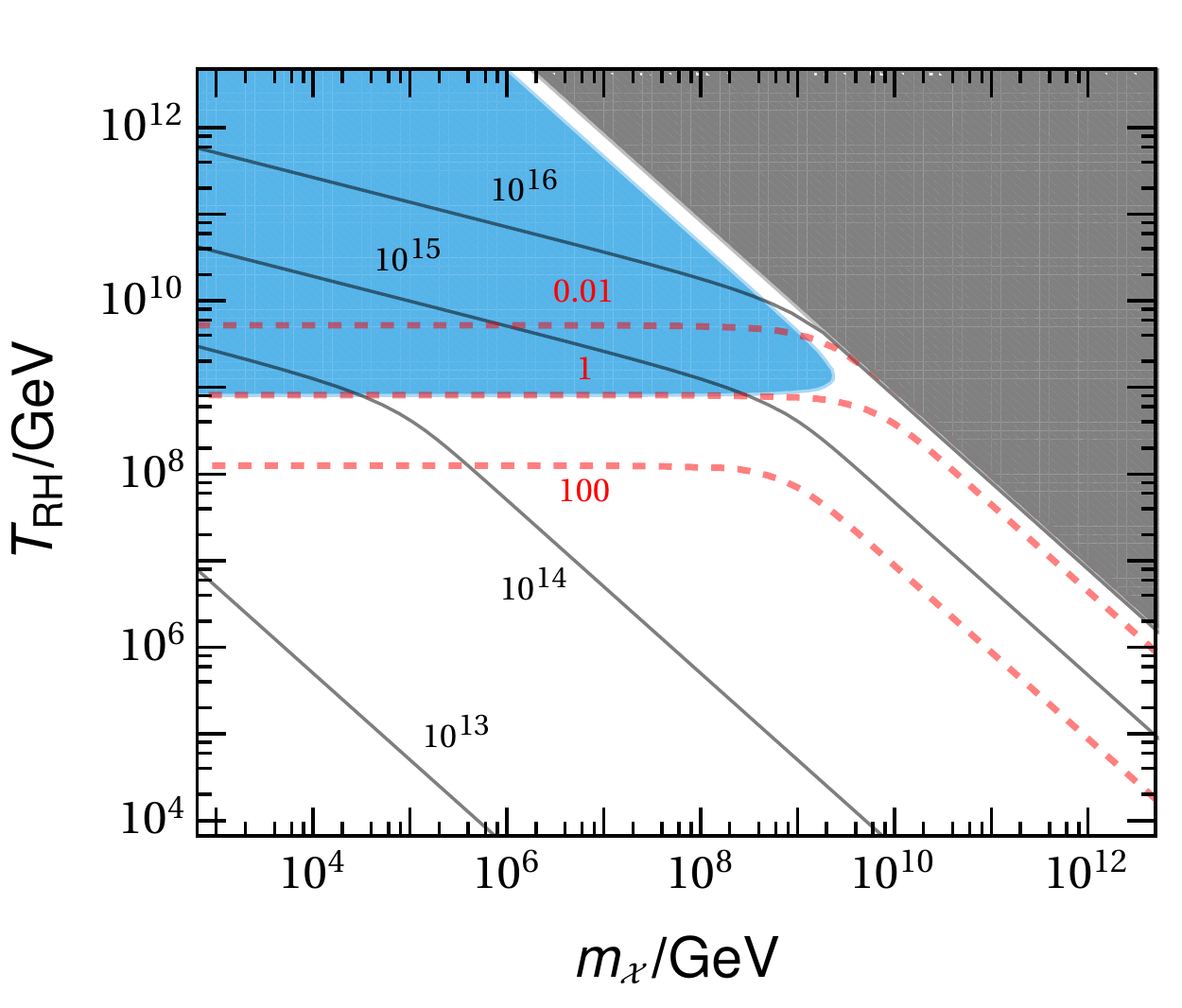}%
	\caption{\label{fig:vportal_compare_th}  Parameter space that realizes the observed dark matter relic density for dimension-6 operators with contribution from thermalization production of dark matter.
    The upper and lower panels are for ${\cal L}_{f\chi}^{V}$ and ${\cal L}_{f{\bf X}}^{V}$, respectively. The left panels and right panels corresponds to $\rho_{{\rm end}}=(10^{16}{\rm GeV})^{4}$
    and $(10^{14}{\rm GeV})^{4}$. The black contours show the required	effective interaction scale, $M/g_{f{\cal X}}$ in GeV; The red line presents the relative importance $n_{{\cal X}}^{{\rm brem}}/(n_{{\cal X}}^{{\rm other}})$ of bremsstrahlung vs other production mechanisms combined for the choice of $M/g_{f{\cal X}}$ that achieves a realistic DM relic. The gray shaded region is where inflaton annihilation through gravity overproduces the DM. The blue shaded areas are dominated by thermalization production.}   
\end{figure}

In Fig.~\ref{fig:vportal_compare_th}, we show the parameter space similar to 
Fig.~\ref{fig:vportal_compare}, but now including the contribution from DM produced during thermalization estimated by  Eq.~\eqref{eq:hard-plasma_prod}. We see that the bremsstrahlung dominant regions are shifted towards lower $\trh$ compare to Fig.~\ref{fig:vportal_compare}. The effect is most
prominent at $m_{\cal X}>\trh$ where the production is suppressed in plasma annihilation but not during the thermalization era due to the energy stored in the
pre-thermalized particles. 
The interpretation of these figure needs some caution. The enhancement factor $(T_{\rm th}/\trh)^{\frac{3}{2}}$ for dimension-6 interactions in Eq.~\eqref{eq:hard-plasma_prod} taken from Ref.~\cite{Harigaya:2019tzu} is only precise up to power counting. It ignored probable numerical factors from the phase space and the number of degrees of freedom.
Actually, in the blue shaded region with $\trh>m_{\cal X}$ in Fig.~\ref{fig:vportal_compare_th}, we get opposite result from Ref.~\cite{Harigaya:2019tzu} regarding the relative importance between plasma annihilation and thermalization production since we have considered various numerical factors in both $n_{\rm th,0}$ and the plasma production according to \cite{Garcia:2018wtq}  while \cite{Harigaya:2019tzu} did not. These factors are enough to modify the result by a few orders of magnitudes. Therefore, at this stage we are unclear whether the $\trh>m_{\cal X}$ parts of the blue shaded regions in Fig.~\ref{fig:vportal_compare_th} is indeed dominated by thermalization or plasma annihilation. What is certain is that deep in the white region the dominant production mechanism must be DM bremsstrahlung. This is because the relative importance between bremsstrahlung and thermalization at $m_{\cal X} < \trh$ scales as $\trh^{-\frac{13}{10}}$. Bremsstrahlung always dominates by orders of magnitudes at low reheating temperature.

\bibliographystyle{utphys}
\bibliography{bremRef}

\end{document}